\documentclass[%
aps,
prd,
amsmath,
amssymb,
reprint,
a4paper,
nofootinbib,
nobibnotes,
superscriptaddress
]{revtex4-2}

\usepackage{newtx}
\usepackage[normalem]{ulem}
\usepackage{dcolumn}
\usepackage[nopmb]{bm}
\usepackage[T1]{fontenc}
\usepackage{pifont}
\usepackage[utf8]{inputenc}
\usepackage[table]{xcolor}
\usepackage{float}
\usepackage{booktabs}
\usepackage{siunitx}
\usepackage{tensor}
\usepackage{leftindex}
\usepackage{mathtools}
\usepackage{array}
\usepackage{orcidlink}
\usepackage{natbib}
\usepackage{graphicx}
\graphicspath{{./figures}}
\usepackage{hyperref}
\hypersetup{pdfpagemode={UseOutlines},
bookmarksopen=true,
bookmarksopenlevel=0,
hypertexnames=false,
breaklinks=true,
colorlinks=true,
citecolor=cyan,
linkcolor=blue,
urlcolor=blue,
unicode
}

\usepackage[capitalize]{cleveref}
\crefname{equation}{}{}
\Crefname{equation}{Eq.}{Eqs.}
\crefname{figure}{Fig.}{Figs.}
\Crefname{figure}{Figure}{Figures}
\crefname{section}{Sec.}{Secs.}
\Crefname{section}{Section}{Sections}

\let\originalleft\left
\let\originalright\right
\renewcommand{\left}{\mathopen{}\mathclose\bgroup\originalleft}
\renewcommand{\right}{\aftergroup\egroup\originalright}

\renewcommand{\arraystretch}{1.2}

\newcommand{\bam}{\textsc{bam}}
\newcommand{\eg}{e.g.,\ }
\newcommand{\ie}{i.e.,\ }
\newcommand{\cf}{cf.\ }

\DeclareSIUnit{\msun}{\mathit{M_\odot}}
\DeclareSIUnit{\gauss}{G}

\begin{document}

\title{Extending the infrastructure of the \bam{} code towards resistive general-relativistic magnetohydrodynamics: tests and first applications}

\author{Matthew Beaudoin~\orcidlink{0000-0001-6597-295X}}
\email{matthew.beaudoin@aei.mpg.de}
\affiliation{Max Planck Institute for Gravitational Physics (Albert Einstein Institute), Am Mühlenberg 1, 14476 Potsdam, Germany}
\affiliation{Institut f\"ur Physik und Astronomie, Universit\"at Potsdam, Haus 28, Karl-Liebknecht-Str. 24/25, 14476 Potsdam, Germany}

\author{Maximiliano Ujevic~\orcidlink{0000-0003-2869-4449}}
\affiliation{Centro de Ci\^encias Naturais e Humanas, Universidade Federal do ABC, 09210-170, Santo Andr{\'e}, 9210-170, SP, Brazil}

\author{Ramon Jaeger}
\affiliation{Institut f\"ur Physik und Astronomie, Universit\"at Potsdam, Haus 28, Karl-Liebknecht-Str. 24/25, 14476 Potsdam, Germany}

\author{Anna Neuweiler~\orcidlink{0000-0003-3205-8373}}
\affiliation{Institut f\"ur Physik und Astronomie, Universit\"at Potsdam, Haus 28, Karl-Liebknecht-Str. 24/25, 14476 Potsdam, Germany}

\author{Henrique Gieg}
\affiliation{Institut f\"ur Physik und Astronomie, Universit\"at Potsdam, Haus 28, Karl-Liebknecht-Str. 24/25, 14476 Potsdam, Germany}

\author{Kenta Kiuchi~\orcidlink{0000-0003-4988-1438}}
\affiliation{Max Planck Institute for Gravitational Physics (Albert Einstein Institute), Am Mühlenberg 1, 14476 Potsdam, Germany}

\author{Tim Dietrich~\orcidlink{0000-0003-2374-307X}}
\affiliation{Institut f\"ur Physik und Astronomie, Universit\"at Potsdam, Haus 28, Karl-Liebknecht-Str. 24/25, 14476 Potsdam, Germany}
\affiliation{Max Planck Institute for Gravitational Physics (Albert Einstein Institute), Am Mühlenberg 1, 14476 Potsdam, Germany}

\date{\today}

\begin{abstract}
Many astrophysical phenomena, including pulsars and short gamma-ray bursts, are
associated with the extremely strong magnetic fields present in neutron stars
and neutron star mergers.
While the ideal magnetohydrodynamic approximation, which assumes infinite
conductivity, provides an excellent description of the neutron-star interior, it
cannot capture non-ideal processes such as Ohmic dissipation and magnetic
reconnection.
To overcome this limitation, we present an extension of the numerical-relativity
code \bam{} incorporating a resistive general-relativistic magnetohydrodynamic
(GRMHD) description.
We validate the new implementation through an extensive suite of
special-relativistic magnetohydrodynamic benchmark tests and by performing
stable simulations of isolated and binary neutron star systems.
For the latter, we investigate the impact of finite conductivity on
magnetic-field amplification, mass ejection, and non-ideal GRMHD effects.
In particular, we find that the component of the electric field parallel to the
magnetic field, which is zero in the ideal case, can reach up to 10\% of the
total electric field strength.
This highlights the potential importance of non-ideal effects for accurately
modeling the long-term evolution of post-merger remnants, particularly in
low-density regions.
Although the present study is restricted to simplified piecewise-polytropic
equations of state, it demonstrates the capabilities of the new resistive GRMHD
framework and paves the way for future investigations employing more realistic
microphysics.
\end{abstract}

\maketitle

\section{Introduction}

Magnetized plasmas play a central role in various high-energy astrophysical
phenomena, including accretion flows onto compact objects in active galactic
nuclei (AGN) or relativistic jets powering gamma-ray bursts (GRBs).
Since the multi-messenger detection of the binary neutron star (BNS) merger
associated with the gravitational-wave (GW) event
GW170817~\cite{LIGOScientific:2017vwq}, along with its electromagnetic
counterparts,
AT2017gfo~\cite{LIGOScientific:2017pwl,Arcavi:2017xiz,Kasen:2017sxr} and
GRB170817A~\cite{Goldstein:2017mmi,Savchenko:2017ffs}, BNS mergers have been
established as engines of short GRBs.
Magnetic fields are expected to be essential for launching relativistic jets
driving such GRBs (\eg\cite{Brandenburg2005,Ruiz:2016rai,Mosta:2020hlh,Kiuchi2024}).

In the last decade, considerable progress has been made in assessing the role of
magnetic fields in BNS mergers and the relevant processes of
magnetohydrodynamics (MHD).
From observations, we know that neutron stars in binary systems typically have
surface magnetic fields of
\(\sim\)$\num{e8}$--$\num{e12}$~\unit{\gauss}~\cite{Lorimer:2008se,Tauris:2017omb}.
While the magnetic field in the inspiral phase is generally considered
negligible for global dynamics, turbulence during the merger triggers MHD
instabilities that rapidly amplify the magnetic field up to
$\sim$\qty{e16}{\gauss}.
Key instabilities include the Kelvin--Helmholtz instability (KHI) that develops
in the shear layer of the colliding stars
(\eg \cite{Price2006,Kiuchi2015,Giacomazzo2015,Aguilera-Miret:2025nts,Gutierrez:2026ngt})
as well as the magnetorotational instability (MRI) driven by differential
rotation in the merger remnant and its surrounding disk,
(\eg\cite{Duez2006,Siegel2013,Kiuchi2024,Aguilera-Miret:2023qih}).
Subsequently, a strong, large-scale magnetic field may be built up in the
remnant by mechanisms such as the MRI-driven $\alpha\Omega$
dynamo~\cite{Kiuchi2024,Most:2023sme} or the Tayler--Spruit
dynamo~\cite{Reboul-Salze:2024jst}, establishing the conditions required to
launch a relativistic jet via the Blandford--Znajek
mechanism~\cite{Blandford1977}.
The magnetic field in BNS mergers thus influences in particular the post-merger
remnant, enhancing angular momentum transport through the MRI and
driving collimated outflows.
Despite this progress, a full picture of these processes is not available and
several key aspects remain uncertain. 

To study the influence of magnetic fields in BNS mergers, we employ
general-relativistic magnetohydrodynamics (GRMHD) to model electrically
conducting fluids in curved spacetime.
In this context are all types of charged particles, \eg electrons and ions,
treated as one continuous fluid, which is a reasonable approximation to describe
the plasma that composes neutron stars.
In BNS simulations, it is common to use the ideal GRMHD approximation
(\eg\cite{Rezzolla2011,Kiuchi2014,Palenzuela2015,Ruiz:2016rai,Kiuchi2018,
Cook2023,Gutierrez:2025gkx,Chabanov2023,Neuweiler2024,Musolino:2024sju};
see \cite{Kiuchi2025} for a recent review).
In this approximation, the magnetic field is effectively ``frozen'' in the fluid
and solely advected by its motion, while the electric field vanishes in the
fluid-comoving frame.
Generally, the electrical conductivity inside neutron stars is expected to be
high enough for ideal GRMHD to be a valid description. 
In cold neutron stars, electrical and thermal transport are primarily determined
by electrons as carriers of charge and heat, which has been extensively studied
in the past (see, \eg\cite{Schmitt:2017efp} for a review).
Progress towards modeling the electrical conductivity of a warm inner and outer
crust of neutron stars has also been made in recent
years~\cite{Harutyunyan2016,Harutyunyan2024}. 
Overall, the magnetic diffusion timescale has been shown to be much longer than
the dynamical timescales of the merger process, permitting the assumption of
``frozen'' magnetic field lines in neutron star matter and for the
applicability of the ideal GRMHD approximation.
Nevertheless, as the temperatures in the remnant increase after the merger and
the densities of the outflowing matter decrease, the conditions to apply ideal GRMHD
might locally break down~\cite{Harutyunyan2018}, and dynamics could be influenced
by non-ideal effects.

In order to systematically assess where the ideal approximation in BNS systems
becomes invalid, numerical simulations using resistive GRMHD are required.
These simulations allow us to perform a quantitative analysis of the impact of
finite conductivity on the system dynamics. 
So far, only a limited number of GRMHD codes account for resistivity
(\eg\cite{Dionysopoulou2013, Palenzuela2013, Cheong2022, Franceschetti2025,
Azizi2025}), and correspondingly few resistive GRMHD simulations of BNS mergers
have been performed~\cite{Palenzuela2013a, Palenzuela2013b, Ponce2014,
Dionysopoulou2015}.
This scarcity is primarily due to conceptual and numerical challenges of
resistive GRMHD, which are avoided in the ideal approximation.
In ideal GRMHD, the electric field can be directly expressed in terms of the
magnetic field and the fluid velocity and thus need not be evolved separately.
In contrast, for resistive GRMHD one must incorporate the evolution equation of
the electric field, which involves a stiff relaxation term from Ampere's law and
complicates numerical time evolution. 
Therefore, robust numerical simulations for resistive GRMHD often rely on
implicit--explicit (IMEX) methods to efficiently handle the stiff equation for
the electric field.
In addition, a consistent formulation of Ohm's law is required to describe the
coupling between electromagnetic fields and fluid variables.

In this work, we extend our current ideal GRMHD framework~\cite{Neuweiler2024}
in the numerical-relativity (NR) code \bam{}~\cite{Brugmann2008,Thierfelder2011}
to include finite electrical conductivities.
We generally follow the approach described by \citet{Palenzuela2009} and
\citet{Dionysopoulou2013}.
This extension allows us to capture Ohmic dissipation and magnetic reconnection,
which are neglected in the ideal approximation.
In particular, the latter might be a relevant mechanism for magnetically
dominated outflows in the polar direction and can produce electromagnetic
flares (\eg\cite{Mbarek:2026xyk}).
In the limit of vanishing conductivity, the electromagnetic fields and matter
decouple and the resistive GRMHD framework reduces to the Maxwell equations in
vacuum, which are clearly distinct from the equations of motion in the ideal GRMHD
approximation.
Thus, the resistive GRMHD framework provides a unified description of regions
with high conductivities inside the stars and with small conductivities outside
the stars.
We validate our new implementation with a series of special- and
general-relativistic tests and present BNS merger simulations performed with
this framework, comparing the results obtained for different conductivities with
those from ideal GRMHD simulations.

The article is structured as follows: \Cref{sec:mathematical} introduces and
outlines the mathematical framework of our resistive GRMHD scheme, while
\cref{sec:numerical} describes the relevant numerical methods; in
\cref{sec:valtests}, numerous validation tests are presented in both flat and
dynamical spacetime; \Cref{sec:bns} presents our BNS merger simulations, and
finally, \cref{sec:conclusion} offers a brief summary of the article.

Throughout this work, we use the $(-,+,+,+)$  metric signature and geometrized
Heaviside--Lorentz units with \mbox{$G = c = \unit{\msun} = \epsilon_0 = 1$}.
We follow the convention where Greek indices represent four-dimensional
spacetime components running from 0 to 3 and Latin indices represent spatial
ones.
The letter $t$ as an index denotes the zeroth or time component of a tensor.

\section{Mathematical framework}\label{sec:mathematical}
\subsection{Spacetime evolution}
We follow the standard 3~+~1 decomposition of the Einstein field equations, in
which the metric has the form
\begin{equation}
    ds^2 = - \alpha^2 dt^2 + \gamma_{ij} \left(dx^i + \beta^i dt\right)
        \left(dx^j + \beta^j dt\right), 
    \label{eq:3+1Decomp}
\end{equation}
where $\alpha$ is the lapse function, $\beta^i$ is the shift vector, and
$\gamma_{ij}$ is the metric of the spacelike hypersurface.
The normal vector to the spacelike hypersurface is denoted as $n^\mu$, which
can be written explicitly as
\begin{equation}
	n^\mu = \left(\alpha^{-1}, -\alpha^{-1}\beta^i\right),\qquad
	n_\mu = \left(-\alpha, 0, 0, 0\right),
\end{equation}
and follows the normalization \(n^a n_a = -1\).
The spatial metric $\gamma_{\mu\nu}$ can be expressed in terms of the spacetime
metric $g_{\mu\nu}$ and $n_\mu$ through the relation $\gamma_{\mu\nu} =
g_{\mu\nu} + n_\mu n_\nu$.

For the general-relativistic simulations, we evolve the spacetime using the Z4c
formulation ~\cite{Bernuzzi2010,Hilditch2013} with $\kappa_1 = 0.02$ and
$\kappa_2 = 0$, along with $1 + \log$ slicing~\cite{Bona1996} and the
$\Gamma$-driver shift condition~\cite{Alcubierre2003}.

\subsection{Relativistic hydrodynamics and the Maxwell equations}
The total energy-momentum tensor for a perfect fluid coupled to an
electromagnetic (EM) field can be written as the sum of two components
\begin{equation}
T^{\mu\nu} = T_\mathrm{fluid}^{\mu\nu} + T_\mathrm{EM}^{\mu\nu},
\end{equation}
given by
\begin{align}
T_\mathrm{fluid}^{\mu\nu} &= \rho h u^\mu u^\nu + p g^{\mu\nu}, \\
T_\mathrm{EM}^{\mu\nu} &= F^{\mu\alpha} F^\nu_{\;\;\alpha} - \frac{1}{4} g^{\mu\nu}
F^{\alpha\beta}F_{\alpha\beta}.
\end{align}
In the above equations, $\rho$ is the rest mass density, $h = 1+\epsilon +
p/\rho$ is the specific enthalpy, $\epsilon$ is the specific internal energy,
and $p$ the hydrodynamical pressure, all measured in the co-moving frame of the
fluid element.
The fluid four-velocity $u^\mu$ is the four-velocity of the fluid element that is measured by an Eulerian
observer.
Note that the electromagnetic tensor $F^{\mu\nu}$ can be related to its dual
$\leftindex^*F^{\mu\nu}$ through the relation
\begin{equation}
    \leftindex^*F^{\mu\nu} = \frac{1}{2} \epsilon^{\mu\nu\alpha\beta} F_{\alpha\beta},
\end{equation}
where $\epsilon^{\mu\nu\alpha\beta} = \eta^{\mu\nu\alpha\beta}/\sqrt{-g}$, with $\eta^{\mu\nu\alpha\beta}$ being the Levi-Civita symbol and $g$ the determinant of the space-time metric $g_{\mu\nu}$.

The complete set of equations, necessary for describing the relativistic motion
of a system, are obtained from the Einstein field equations, the energy-momentum
conservation equations
\begin{equation}
    \nabla_\mu T^{\mu\nu} = 0,
    \label{eq:energymom}
\end{equation}
and the Maxwell equations, written in covariant form as
\begin{equation}
    \nabla_\nu F^{\mu\nu} = I^\mu, \qquad \nabla_\nu\leftindex^*F^{\mu\nu} = 0,
    \label{eq:maxwell} 
\end{equation}
where $I^\mu$ is the four-current, together with the conservation of rest mass
and the conservation of electric charge
\begin{equation}
    \nabla_\mu \left(\rho u^\mu\right) = 0, \qquad \nabla_\mu I^\mu = 0, \label{eq:masscurrent}
\end{equation}
respectively.
Moreover, with the use of the normal vector to the spatial hypersurface,
$n^\mu$, we can express the fluid four-velocity $u^\mu$, the four-current
$I^\mu$, and the electromagnetic tensors $F^{\mu\nu}$ and
$\leftindex^*F^{\mu\nu}$ as
\begin{align}
    u^\mu &= W \left(n^\mu + v^\mu \right),\\
    I^\mu &= q n^\mu + J^\mu, \label{eq:four-current}\\
    F^{\mu\nu} &= n^\mu E^\nu - n^\nu E^\mu + \epsilon^{\mu\nu\alpha\beta}
        B_\alpha n_\beta,\\
    \leftindex^*F^{\mu\nu} &= n^\mu B^\nu - n^\nu B^\mu -
        \epsilon^{\mu\nu\alpha\beta} E_\alpha n_\beta,
\end{align}
where $v^\mu$ corresponds to the three-dimensional velocity measured by an
Eulerian (normal) observer, $W = \alpha u^t = 1/ \sqrt{1 - v_i v^i}$ is the
Lorentz factor, $q$ is the electric charge, $E^\mu$ and $B^\mu$ are the purely
spatial electric and magnetic fields ($n_\mu E^\mu = n_\mu B^\mu = 0$), and
$J^\mu$ is the purely spatial current ($n_\mu J^\mu = 0$).

\subsection{Divergence cleaning}
Numerically, solving the Maxwell equations \cref{eq:maxwell} presents several
difficulties.
One of them is how to prevent the emergence of magnetic monopoles (violations
of the solenoidal constraint $\nabla_i B^i = 0$) during the simulation that
arise as a result of numerical imprecision.
This problem has been addressed in the past using different techniques, such as
constrained transport schemes~\cite{Evans1988, Shankar2023, Kiuchi2022,
Cook2023}, the vector potential method~\cite{Etienne2010, Cipolletta2020}, and
the divergence cleaning method~\cite{Dedner2002, Liebling2010, Penner2011a,
Moesta2014, Deppe2021, Palenzuela2018}.
Because of the structure of \bam{}, the divergence cleaning method is the most
straightforward to implement~\cite{Neuweiler2024}.
In this approach, new scalar fields $\phi$ and $\psi$ are introduced into the
Maxwell equations \cref{eq:maxwell}.
This was first proposed in the context of ideal MHD by \cite{Dedner2002} and
later by \cite{Komissarov2007} in the context of resistive MHD.
The new equations read as
\begin{align}
\nabla_\nu \left( F^{\mu\nu}-g^{\mu\nu}\psi \right) &= I^\mu -
\kappa_\psi n^\mu \psi, \label{maxwellext1} \\
\nabla_\nu \left(\leftindex^*F^{\mu\nu}-g^{\mu\nu} \phi \right) &= -\kappa_\phi n^\mu \phi.
\label{eq:maxwellext2}
\end{align}
Projecting the above equations parallel to $n_\mu$, we obtain
\begin{align}
\left(\partial_t - \cal{L}_\beta \right)\psi + \alpha \nabla_i E^i &= -\alpha
    \kappa_\psi \psi + \alpha q, \label{eq:psicleaning} \\
\left(\partial_t - \cal{L}_\beta \right)\phi + \alpha \nabla_i B^i &= - \alpha
    \kappa_\phi \phi \label{eq:phicleaning},
\end{align}
where $\cal{L}_\beta$ is the Lie derivative along the shift vector $\beta^i$.
Here, we see that the scalar fields $\phi$ and $\psi$ measure the deviation
from the electromagnetic constrained solution.
The fields $\phi$ and $\psi$ satisfy a damped wave equation propagating at the
speed of light decaying exponentially to the trivial solution $\phi = \psi = 0$
over a timescale $\sim$$1/\kappa_\phi$ and $\sim$$1/\kappa_\psi$, respectively.
It is clear from \cref{eq:psicleaning} that the new field $\psi$ will
drive the solution towards the condition $\nabla_i E^i = q$.
We see on the other hand from \cref{eq:phicleaning} that $\phi$ will drive the
solution to the zero-divergence condition $\nabla_i B^i = 0$.
Note that in the limit $\phi,\,\psi \rightarrow 0$, the augmented Maxwell
equations are reduced to the standard Maxwell equations (\ref{eq:maxwell}).

\subsection{Ohm's law}
The general covariant form of the electric current $I^\alpha$
satisfies~\cite{Tsamparlis2010}
\begin{equation}
    h_{\mu\nu} I^\nu = \sigma_{\mu\nu} F^\nu_{\;\;\alpha} u^\alpha,
\end{equation}
where $\sigma_{\mu\nu}$ is the electric conductivity tensor, and $h_{\mu\nu} =
g_{\mu\nu} + u_\mu u_\nu$ is the projector operator orthogonal to $u^\mu$.
In this work, we make the approximation that the medium is isotropically
and homogeneously conductive.
Therefore, we can write $\sigma_{\mu\nu} = \sigma g_{\mu\nu}$.
With these considerations, Ohm's law can be written for a fluid element as
\begin{equation}
    J^i = q v^i + W \sigma \left[E^i + \epsilon^{ijk} v_j B_k - \left(v_k
    E^k\right)v^i \right], \label{eq:ohmlaw}
\end{equation}
similarly to~\cite{Palenzuela2009, Dionysopoulou2013}.
The first term on the right-hand side of \cref{eq:ohmlaw} corresponds to the
convection current, which depends on the motion of the charged medium.
The second term corresponds to the conduction current, which depends on the
electrical conductivity $\sigma$ of the plasma and the electromagnetic fields.

In neutron stars, it is expected that the electrical conductivity will attain
values as high as \qty{e25}{s^{-1}} in the crust and \qty{e29}{s^{-1}} in the
core~\cite{Harutyunyan:2016rxm, Harutyunyan2018, Pons2019, Harutyunyan:2023ooz, Petrosyan:2025nrl}.
The presence of high values of $\sigma$ leads to stiff terms in the evolution
equations that are, in practice, impossible to handle with explicit
Runge--Kutta methods.
This forces us to use more elaborate evolution schemes, such as
implicit--explicit Runge--Kutta (IMEX) methods~\cite{Ascher1997, Pareschi2005,
Izquierdo2023}.
During this work we use conductivities as high as $\sigma=10^6$, which already
show results similar to the ideal case.%
\footnote{We note that a factor $\qty{1.62e4}{s^{-1}}$ is needed to convert the
conductivity to cgs units.}
In \cref{sec:imex}, we discuss our IMEX implementation.

\subsection{Evolution equations}
When expressing the dynamical equations \cref{eq:energymom},
\cref{eq:masscurrent}, \cref{maxwellext1} and \cref{eq:maxwellext2} in flux-conserved form, it is
useful to define the following projections of the energy-momentum tensor:
\begin{align}
    \label{eq:projU}
    U = n_\mu n_\nu T^{\mu\nu} &\equiv \rho h W^2 - p + \frac{1}{2}
        \left(E^2 + B^2\right), \\
    \label{eq:projS_i}
    S_i = -n^\mu \tensor{\gamma}{^\nu_i} T_{\mu\nu} &\equiv \rho h W^2 v_i +
        \epsilon_{ijk} E^j B^k, \\
    %
    \label{eq:projS_ij}
    \begin{split}
        S_{ij} = \tensor{\gamma}{^\mu_i} \tensor{\gamma}{^\nu_j} T_{\mu\nu}
    &\equiv \rho h W^2 v_i v_j + \gamma_{ij}p - E_i E_j - B_i B_j \\&\qquad +
    \frac{1}{2} \gamma_{ij} \left(E^2 + B^2\right),
    \end{split}
\end{align}
where $U$ is the total energy density, $S_i$ is the momentum density, and
$S^{ij}$ is the spatial stress, all of them as measured by an Eulerian observer.
With these definitions, the complete set of evolution equations can be written
as
\begin{widetext}
    \begin{align}
    \partial_t \left(\sqrt{\gamma}D\right) &+ \partial_i \left[\sqrt{\gamma}D
        \left(\alpha v^i - \beta^i\right) \right] = 0, \label{eq:D} \\
    \partial_t \left( \sqrt{\gamma} \tau \right) &+ \partial_i \left\{
        \sqrt{\gamma} \left[ \alpha \left(S^i - v^iD\right) - \beta^i \tau
        \right] \right\} = \sqrt{\gamma} \left( \alpha S^{ij} K_{ij} - S^i
        \partial_i \alpha \right), \\
    \partial_t \left( \sqrt{\gamma} S_i\right) &+ \partial_k \left[ \sqrt{\gamma}
        \left( \alpha S_i^{\;k} - S_i \beta^k \right) \right]  =\sqrt{\gamma}
        \left[ \frac{\alpha}{2} S^{km} \partial_i \gamma_{km} + S_k \partial_i
        \beta^k - (\tau + D) \partial_i \alpha \right], \\
    \begin{split}
    \partial_t \left( {\sqrt{\gamma}B^i} \right) &+ \partial_j \left[ \sqrt{\gamma}
        \left( -\beta^j B^i + \alpha \epsilon^{ijk} E_k + \alpha \gamma^{ij}
        \phi \right) \right] \\ &\qquad\qquad = -\sqrt{\gamma} B^j \partial_j \beta^i +
        \sqrt{\gamma} \phi \left[ \gamma^{ij} \partial_j \alpha + \alpha \left(
        \frac{1}{2} \gamma^{ij} \gamma^{kl} \partial_j \gamma_{kl} - \gamma^{ki}
        \gamma^{lj} \partial_j \gamma_{kl} \right) \right], \label{eq:magnetic}
    \end{split}\\
    \begin{split}
    \partial_t \left( {\sqrt{\gamma}E^i} \right) &+ \partial_j \left[ \sqrt{\gamma}
    \left( -\beta^j E^i - \alpha \epsilon^{ijk} B_k + \alpha \gamma^{ij} \psi
    \right) \right] \\ &\qquad\qquad = -\sqrt{\gamma} E^j \partial_j \beta^i + \sqrt{\gamma} \psi
    \left[ \gamma^{ij} \partial_j \alpha + \alpha \left( \frac{1}{2} \gamma^{ij}
    \gamma^{kl} \partial_j \gamma_{kl} - \gamma^{ki} \gamma^{lj} \partial_j
    \gamma_{kl} \right) \right] - \alpha \sqrt{\gamma}J^i, \label{eq:electric}
    \end{split}\\
    \partial_t \phi &+ \partial_i \left( \alpha B^i - \beta^i \phi \right) =-
        \phi \partial_i \beta^i + B^i \partial_i \alpha - \frac{\alpha}{2} B^i
        \gamma^{lm} \partial_i \gamma_{lm} - \alpha \kappa_\phi \phi, \\
    \partial_t \psi &+ \partial_i \left( \alpha E^i - \beta^i \psi \right) =-
        \psi \partial_i \beta^i + E^i \partial_i \alpha - \frac{\alpha}{2} E^i
        \gamma^{lm} \partial_i \gamma_{lm} +\alpha q - \alpha \kappa_\psi \psi,
        \\
    \partial_t \left( \sqrt{\gamma} q\right) &+ \partial_i \left[ \sqrt{\gamma}
        \left( \alpha J^i - q\beta^i \right) \right] = 0, \label{eq:q}
    \end{align}
\end{widetext}
where $D = \rho W$ and $\tau = U - D$.
The fields $\{\sqrt{\gamma}D$, $\sqrt{\gamma}\tau$, $\sqrt{\gamma}S_i$,
$\sqrt{\gamma}B^i$, $\sqrt{\gamma}E^i$, $\phi$, $\psi$, $\sqrt{\gamma}q\}$ are
called the conserved fields.
The conserved quantities $D$, $\tau$~\cref{eq:projU}, and
$S_i$~\cref{eq:projS_i} can be written in terms of the primitive variables
$\{\rho$, $\epsilon$, $v^i$, $p$, $E^i$, $B^i\}$.
$K^{ij}$ is the extrinsic curvature and $\gamma$ is the determinant of the
spatial metric $\gamma_{ij}$.
Note that in \cref{eq:magnetic} and \cref{eq:electric}, following
\citet{Moesta2014}, we have made the substitutions
\begin{align}
-\alpha &\sqrt{\gamma} \gamma^{ij} \partial_j \phi = -\partial_j \left( \alpha \sqrt{\gamma} \gamma^{ij} \phi \right) \nonumber \\
&+ \sqrt{\gamma} \phi \left[ \gamma^{ij} \partial_j \alpha + \alpha \left( \frac{1}{2} \gamma^{ij} \gamma^{kl} \partial_j \gamma_{kl} - \gamma^{ki} \gamma^{lj} \partial_j \gamma_{kl} \right) \right], \\ 
-\alpha &\sqrt{\gamma} \gamma^{ij} \partial_j \psi = -\partial_j \left( \alpha \sqrt{\gamma} \gamma^{ij} \psi \right) \nonumber \\
&+ \sqrt{\gamma} \psi \left[ \gamma^{ij} \partial_j \alpha + \alpha \left( \frac{1}{2} \gamma^{ij} \gamma^{kl} \partial_j \gamma_{kl} - \gamma^{ki} \gamma^{lj} \partial_j \gamma_{kl} \right) \right],
\end{align}
in contrast to \cite{Dionysopoulou2013} but similar to \cite{Azizi2025}, in
order to avoid spatial derivatives of the auxiliary fields in the source terms.

\section{Numerical methods}\label{sec:numerical}
We use the \bam{} code \cite{Brugmann1996, Brugmann1999, Brugmann2008,
Thierfelder2011, Dietrich2015, Bernuzzi2016} for our NR
simulations.
We refer the reader to \citet{Neuweiler2024} for a description of the
ideal GRMHD implementation, including a detailed description of the grid
structure and numerical methods.
In \bam{}, the equations of 3~+~1 NR are solved with a method
of lines approach.
Finite-difference methods are employed for spatial derivatives and the equations
are evolved forward in time via a Runge--Kutta integrator.

The numerical domain is represented by a hierarchical series of nested
three-dimensional Cartesian grid boxes called refinement levels.
There are $L$ levels indexed by the integer parameter $\ell = 0,1,2,\ldots,L-1$.
We use a binary refinement scheme; in other words, the grid spacing $h_\ell$ on
level $\ell$ is given by $h_\ell = h_0 / 2^\ell$.
The refinement scheme is entirely nested, meaning that the
coordinate extent of any grid at level $\ell \geq 1$ is completely covered by the
grid at level $\ell - 1$.
For inner refinement levels with $\ell \geq \ell_\mathrm{move}$, the Cartesian
boxes can move and adjust dynamically during the evolution to track the compact
objects during inspiral.
To handle boundaries, all levels with $\ell \geq 1$ contain a six-point buffer
region where prolongation of the data from the coarser level takes place;
conversely, data is restricted from the finer level to the coarser one in
all regions of overlap.
In this work, we use sixth-order Lagrangian polynomial interpolation for the
geometric variables and linear interpolation for matter
variables;~\cf\cite{Dietrich2015}.

Time evolution is handled according to the Berger--Oliger scheme~\cite{Berger1984}
with an additional Berger--Colella correction step~\cite{Berger1989}, which
reinforces conservation laws by correcting fluxes across refinement
boundaries~\cite{Dietrich2015}.
The Courant factor is the same for each level, so that if level $\ell$ takes a
timestep $\Delta t$, then level $\ell + 1$ takes two timesteps with $\Delta t /
2$.
As is common practice, we define a minimum level $\ell_{\mathrm{BO}}$ at which
time refinement will be performed; the levels coarser than this threshold will
all take the same timestep as level $\ell = \ell_{\mathrm{BO}}$.

\subsection{Implicit--explicit time integration}\label{sec:imex}
Since the conductivity is expected to undergo large spatial variations in
neutron star simulations, from the highly conductive interior to the tenuous
exterior, dynamical timescales can vary greatly across the numerical domain.
We must therefore mathematically regard the problem as a hyperbolic system with
relaxation terms, which requires special care to capture the dynamics in a
stable and accurate manner.
A prototypical hyperbolic system of partial differential equations with a
relaxation term can be written
\begin{equation}
	\label{eq:hyperbolic-relaxation}
	\partial_t\bm{U} = F\left(\bm{U}\right) +
		\frac{1}{\varepsilon}R\left(\bm{U}\right),
\end{equation}
where \(\bm{U}\) is the state vector of all evolved (conserved) variables,
\(F(\bm{U})\) and \(R(\bm{U})\) are right-hand side (RHS) functions that depend on \(\bm{U}\) and
its spatial derivatives, and \(\varepsilon = 1/\sigma > 0\) is the relaxation time.
In the stiff limit ($\varepsilon \rightarrow 0$), a stable explicit evolution
can only be achieved with a timestep size $\Delta t \leq \varepsilon$.
Naturally, this is unfeasible for explicit integration of conductive plasmas.
Implicit methods can handle relaxation terms, as they are not confined by this
debilitating limit on $\Delta t$, but they are much more computationally
expensive.

Due to the presence of the stiff term in (\ref{eq:electric}), \ie the one
containing the conductivity $\sigma$ inside $J^i$, we use an implicit--explicit
(IMEX) Runge--Kutta (RK) method~\cite{Ascher1997, Pareschi2005, Boscarino2007,
Izquierdo2023}.
Such a method has been employed by many groups in the context of resistive
relativistic MHD (\eg\cite{Palenzuela2009, Dionysopoulou2013, Miranda-Aranguren2018,
Ripperda2019a, Ripperda2019b, Wright2020, Cheong2022, Mignone2024,
Franceschetti2025, Azizi2025}), and in this section we summarize it.
The scheme consists of applying an implicit evolution for the stiff terms and an
explicit evolution for the non-stiff terms, such that
\cref{eq:hyperbolic-relaxation} can be integrated as
\begin{align}
    \bm{U}^{(i)} &= \bm{U}^n
    + \Delta t \sum_{j=1}^{i-1}
        \tilde{a}_{ij}F\left[\bm{U}^{(j)}\right]
    + \Delta t \sum_{j=1}^{i}
        a_{ij}\frac{1}{\epsilon}R\left[\bm{U}^{(j)}\right],\\
	\label{eq:imex-update}
	\bm{U}^{n + 1} &= \bm{U}^n
		+ \Delta t \sum_{i=1}^{q}
			\tilde{\omega}_{i}F\left[\bm{U}^{(i)}\right]
		+ \Delta t \sum_{i=1}^{q}
			\omega_i \frac{1}{\epsilon}R\left[\bm{U}^{(i)}\right] .
\end{align}
In the above equations, $\bm{U}^{(i)}$ are the intermediate substep values of
the $q$-stage RK method chosen.
The matrices $\tilde{a}_{ij}$, $a_{ij}$, and the vectors $\tilde{\omega}_i$,
$\omega_i$ are calculated such that the resulting scheme is diagonally implicit in
$R(\bm{U})$ and explicit in $F(\bm{U})$.
These quantities are usually expressed in the form of Butcher tableaux.
In this work, we use the classical fourth-order RK for the explicit part and a
second-order $L$-stable method for the implicit part, corresponding to the
``IMEX42L'' scheme of \citet{Izquierdo2023}; see the reference for the
explicit form of the Butcher tableaux.

We split the state vector of evolved fields \(\bm{U}\) into two sets of
variables \(\{\bm{V}, \bm{W}\}\) depending on the presence of a relaxation
(stiff) term in their evolution equation.
In our case, only the equation of the electric field contains a relaxation term,
so we split the fields according to \(\bm{V} = \{\sqrt{\gamma}E^i\}\) and
\(\bm{W} =
\{
	\sqrt{\gamma}D,
	\sqrt{\gamma}\tau,
	\sqrt{\gamma}S^i,
	\sqrt{\gamma}B^i,
	\phi,
	\psi
\}\).
We then separate \cref{eq:hyperbolic-relaxation} into individual equations
\begin{align}
	\partial_t \bm{W} &= F_W\left(\bm{V}, \bm{W}\right),\\
	\partial_t \bm{V} &= F_V\left(\bm{V}, \bm{W}\right)
		+ \frac{1}{\epsilon\left(\bm{W}\right)} R_V\left(\bm{V}, \bm{W}\right).
\end{align}
The dependence of the relaxation parameter \(\epsilon\) on \(\bm{W}\) is
arbitrary.
In this arrangement, however, it must not depend on the stiff \(\bm{V}\)
fields.

Each RK stage $\bm{U}^{(i)}$ in \cref{eq:imex-update} can be divided into two
further substeps.
In the first substep, the explicit intermediate values $\{\bm{V}^*, \bm{W}^*\}$
are calculated according to
\begin{align}
	\label{eq:implicitWdef}
	\bm{W}^* &= \bm{W}^n
		+ \Delta t \sum_{j=1}^{i-1}
			\tilde{a}_{ij}F_W\left[\bm{U}^{(j)}\right],\\
	\label{eq:implicitVdef}
	\bm{V}^* &= \bm{V}^n
		+ \Delta t \sum_{j=1}^{i-1}
			\tilde{a}_{ij}F_V\left[\bm{U}^{(j)}\right]
		+ \Delta t \sum_{j=1}^{i-1}
			a_{ij}\frac{1}{\epsilon\left[\bm{W}^{(j)}\right]}
			R_V\left[\bm{U}^{(j)}\right].
\end{align}
Then, in the second substep, one obtains the full value $\bm{U}^{(i)}$ by solving the
equations
\begin{align}
	\label{eq:implicitWupdate}
	\bm{W}^{(i)} &= \bm{W}^*,\\
	\label{eq:implicitVupdate}
	\bm{V}^{(i)} &= \bm{V}^*
		+ a_{ii}\frac{\Delta t}{\epsilon\left[\bm{W}^{(i)}\right]}
			R_V\left[\bm{V}^{(i)}, \bm{W}^{(i)}\right].
\end{align}
It is evident that $\bm{W}^{(i)}$ is trivially updated and that \cref{eq:implicitVupdate}
is an implicit equation for $\bm{V}^{(i)}$.
Inserting the stiff part of \cref{eq:electric} and $1/\varepsilon[\bm{W}^{(i)}]
= \sigma$, for our choice of Ohm's law \cref{eq:ohmlaw},
\cref{eq:implicitVupdate} takes the form (dropping the $(i)$ superscript for
clarity)
\begin{align}
    E^i = E^{i*} - \alpha \Delta t a_{ii} W \sigma \left[E^i + \epsilon^{ijk} v_j B_k - \left(v_k
    E^k\right)v^i \right], 
\end{align}
which is a linear system that can be inverted to give the direct update
\begin{align}
    E^i = \mathbb{M}^{-1} \left(E^{i*} - \bar{\sigma} \epsilon^{ijk} v_j B_k \right), \label{eq:direct}
\end{align}
where $\bar{\sigma} = \alpha\Delta t a_{ii} W \sigma$ and
\begin{equation}
	\mathbb{M} =
	\begin{bmatrix}
		1 + \bar{\sigma}\left(1 - v_x v^x\right)
			& -\bar{\sigma}\left(v_y v^x\right)
			& -\bar{\sigma}\left(v_z v^x\right)  \\
		-\bar{\sigma}\left(v_x v^y\right)
			& 1 + \bar{\sigma}\left(1 - v_y v^y\right)
			& -\bar{\sigma}\left(v_z v^y\right)  \\
		-\bar{\sigma}\left(v_x v^z\right)
			& -\bar{\sigma}\left(v_y v^z\right)
			& 1 + \bar{\sigma}\left(1 - v_z v^z\right) \\
	\end{bmatrix}.
\end{equation}

With the description of the methods presented above, we are in the position to explain our time integration procedure, which is similar to the one presented in \cite{Palenzuela2013, Dionysopoulou2013}.

(1) We start with the first RK stage $\bm{U}^{(1)}$.
In the IMEX42L scheme, this substep contains no implicit calculation, \ie
$\bm{U}^{(1)} = \bm{U}^n$.
Therefore, the electromagnetic fields $E^i$ and $B^i$ are calculated from the
previous iteration just by dividing the conserved electromagnetic fields by
$\sqrt{\gamma}$.
To obtain the hydrodynamic primitives, we first subtract the
contribution of the electromagnetic field from the conserved quantities
\begin{align}
    \bar{\tau} &= \tau - \frac{1}{2}\left(E^2 + B^2\right),\\
    \bar{S}_i &= S_i - \epsilon_{ijk} E^j B^k,
\end{align}
from which we proceed with the conservative-to-primitive inversion of \citet{Galeazzi2013}.

(2) For all remaining RK substeps an implicit solve is required, and we have the
situation described in \cref{eq:implicitWdef}--\cref{eq:implicitVupdate}.
In this case, the explicit conserved quantities $\bm{W}^*$ have already been
calculated, and we use an iterative procedure to find the primitives, which are
required for the implicit solve.
We numerically search for roots of the function
\begin{align}
    f(p) = p_\mathrm{EOS}(\rho,\epsilon) - p , \label{eq:proot}
\end{align}
where $p_\mathrm{EOS}(\rho,\epsilon)$ is given by the chosen equation of state (EOS).
A common procedure is to use the Newton--Raphson method, where the value of $p$
in iteration $m+1$ is updated using values from iteration $m$ via
\begin{align}
    p_{m+1} = p_m - \frac{f(p_m)}{f'(p_m)}, \label{eq:pm+1}
\end{align}
where $f'(p) = v^2 c^2_s - 1$ and $c_s$ is the local sound speed of the fluid.
We begin with the previously calculated values of $p$, $v^i$, $E^{i*}$, and
$B^i$, from which we obtain, via \cref{eq:direct}, the new value of $E^i$.
Then, with the updated electromagnetic fields, we calculate $\bar{\tau}$ and
$\bar{S_i}$ as in step 1, and obtain in the following order
\begin{align}
    v_i &= \frac{\bar{S}_i}{\bar{\tau}+D+p}, \label{eq:viteration} \\
    W &= \frac{1}{\sqrt{1-v_iv^i}}, \\
    \rho &= \frac{D}{W}, \\
    \epsilon &= \frac{\bar{\tau}+\left(1-W\right)D+\left(1-W^2\right)p}{WD}.
\end{align}
With the values of $\epsilon$, $\rho$, and $p$, together with \cref{eq:proot},
we obtain the new value $p_{m+1}$ using \cref{eq:pm+1}.
These new values of $p$ and $v_i$ (\ref{eq:viteration}) are then used as the initial guess for the next iteration.
We continue this operation until a relative error of \num{e-10} is obtained for
the pressure and individual components of the electric field, which occurs after
about 10 iterations for most grid points.
In some 2D and 3D applications that have a strong magnetic field in only one
direction, we consider instead, as an indicator of convergence, the relative
error of the magnitude of the electric field, which must also be less than
\num{e-10}.

As in \cite{Dionysopoulou2013}, we decided not to evolve the electric charge $q$ using (\ref{eq:q}), but to calculate it directly from Gauss's law
\begin{equation}
    q = \nabla_i E^i = \frac{1}{\sqrt{\gamma}} \partial_i \left(\sqrt{\gamma}E^i\right),
\end{equation}
which avoids the problem that may appear when large gradients of the current are present.

\subsection{Atmosphere}\label{sec:atm}
Simulating the interface between the edge of a neutron star and the vacuum (or
extremely tenuous) region outside of it poses numerical difficulties, even in
the simpler Newtonian case~\cite{Toro2009}.
Primitive recovery techniques can become singular for $\rho = 0$, and so special
care would have to be taken for these regions, even if it were feasible to
handle them using high-resolution shock-capturing (HRSC) methods.
To address this, \bam{} applies a cold and static ``atmosphere'', as is
prevalent in the literature (\eg\cite{Duez2003, Baiotti2005, Giacomazzo2007,
Moesta2014, Porth2017, Qian2017, Shankar2023, Cook2025}).
In \bam{} we approximate vacuum regions by setting a floor value for $\rho$ many
orders of magnitude below the maximum density $\rho_c$~\cite{Thierfelder2011}.
The atmosphere density value is computed from the initial data at the start of
the simulation by $\rho_\mathrm{atm} = f_\mathrm{atm} \rho_c$, where
$f_\mathrm{atm}$ is the chosen ``atmosphere fraction''.
During the simulation, if any point has a density less than a threshold value
$\rho_\mathrm{thr} = f_\mathrm{thr} \rho_\mathrm{atm}$, we manually set the
density of that point to $\rho_\mathrm{atm}$.
This prevents us from being sensitive to any fluctuations around the density
floor.
In the atmosphere, we set the pressure $p$ and energy density $\epsilon$
according to the zero-temperature part of the EOS (a ``cold'' atmosphere).
We also force a ``static'' atmosphere by setting the three-velocity $v^i = 0$.
Atmosphere-setting occurs during each Runge--Kutta substep, so in practice,
atmosphere points do not experience hydrodynamic evolution.

Some (single and binary) neutron star simulations in ideal GRMHD use an
unmagnetized atmosphere (\eg\cite{Del_Zanna2003, Shibata2005}), while others
evolve the EM fields normally in the atmosphere (\eg\cite{Giacomazzo2007,
Neuweiler2024}).
Since one of the primary incentives for the resistive framework as presented
here is a unified electromagnetic treatment for the neutron star interior and
 exterior, we naturally let the EM fields evolve in the atmosphere.
In the single and binary neutron star simulations, we choose a zero-conductivity
atmosphere in which the EM fields are governed by the Maxwell equations in
vacuum (local charge density would also have an effect would we not set $v^i =
0$).

\subsection{Riemann solver and characteristic speeds}\label{sec:riemann}
The numerical flux at the cell interface is given by an approximate solution to
the Riemann problem.
We use the single-speed local Lax--Friedrichs (LLF) flux
formula.
The overall characteristic speed is computed by taking the
fastest left- and right-moving speeds as computed for the left and right state,
such that
\begin{equation}
	\lambda_\pm  = \max\left\{0,
		\pm\lambda_\pm^L,
		\pm\lambda_\pm^R\right\},\qquad
	\lambda = \max\left\{\lambda_+, \lambda_-\right\}.
\end{equation}
Here, $\lambda_\pm^{L}$ and $\lambda_\pm^{R}$ are the
characteristic wave speeds at the cell interface determined from the $L$ and $R$
state, respectively.
This simplified approach to the Riemann problem is advantageous as one need not
know the characteristic eigenstructure nor all corresponding eigenvalues
\cite{Del_Zanna2003}.
The latter are known for ideal GRMHD, but exact or approximate forms of the
eigenvalues of the resistive GRMHD system (outside of limiting cases) are not
(\eg\cite{Mignone2018, Miranda-Aranguren2018, Schoepe2018}).

The 3~+~1 nature of our system means that the eigenvalues in the $i$-direction
correspond to characteristic waves of the form
(\eg\cite{Anile1989, Del_Zanna2007, Ripperda2019a})
\begin{equation}
	\lambda^i_\pm = \alpha\tilde{\lambda}^i_\pm - \beta^i
\end{equation}
where $\tilde{\lambda}^i$ is the characteristic speed in the locally flat frame.
For the sake of simplicity, and as is usually done in resistive GRMHD
simulations (\eg\cite{Bucciantini2013, Dionysopoulou2015, Ripperda2019a, Azizi2025}),
we take the characteristics to be in the limit of maximum diffusivity.
That means that the maximum wave speeds entering into the Riemann solver are the
speed of light, which upon transforming to the Eulerian frame \cite{Pons1998}
gives $\tilde{\lambda}^i_\pm = \pm\sqrt{\gamma^{ii}}$~\cite{Del_Zanna2007}.
This is reasonably justified, as in the resistive case the maximum wavespeed is
not limited by the fast magnetosonic mode and can approach the speed of light
(which is also the characteristic speed of the constraint damping variables
$\phi$ and $\psi$) independently of the value of the magnetic
field~\cite{Bucciantini2013}.
However, this can lead to some smearing of results in the ideal MHD limit,
where sharp field discontinuities may arise, but is a less unsuitable
approximation for properly resistive regimes as more diffusion is expected
overall.

\section{Validation tests}\label{sec:valtests}
To validate our implementation, we present in the following sections a
suite of special-relativistic tests, where, unless otherwise stated, we employ
the weighted essentially non-oscillatory (WENOZ) reconstruction
method~\cite{Jiang1996,Borges2008,Castro2011},
the local Lax–Friedrichs (LLF) Riemann solver, a Courant factor of 0.25, and an
ideal gas EOS with $\Gamma = 2$ for the one-dimensional tests and $\Gamma = 4/3$
for the two- and three-dimensional tests.
General-relativistic simulations of a single magnetized neutron star are also
presented, for which a piecewise-polytropic fit of the SLy EOS is used.

\subsection{One-dimensional shock tube}\label{sec:shocktube}
One of the most fundamental and prevalent tests in (magneto)hydrodynamic
simulations is the ``shock tube'', which usually consists of a one-dimensional
domain (the tube) with an initial discontinuity in one or more of the fluid
variables.
In purely hydrodynamic simulations, these are typically the pressure, rest-mass
density, fluid velocity, or internal energy; one may also introduce
discontinuities in the electromagnetic fields when using MHD codes.
To examine the code's ability to successfully capture MHD shocks, we perform a
shock tube test as proposed by~\citet{Brio1988} and later modified
by~\citet{Giacomazzo2006}.

We evolve the shock tube between two limiting regimes with all tests sharing a
single set of initial conditions.
The first regime is an approximate recovery of ideal MHD ($\sigma_0 \rightarrow
\infty$) and the second is the zero-conductivity limit ($\sigma_0 \rightarrow
0$).
With our choice of the isotropic scalar Ohm's law, the latter corresponds to the
electrovacuum limit.
We explore these limits via the conductivity $\sigma$, with one variation of the
test using a constant uniform conductivity $\sigma = \sigma_0$ and the other
parameterizing $\sigma$ according to a power-law function of the conserved
rest-mass density to examine how the code handles spatial variations in the
conductivity. 

The tests are run on a one-dimensional grid with $x \in [0.5, 0.5]$.
There is an initial discontinuity at $x = 0$ with a left (L) and right (R)
state given by
\begin{equation}
	\begin{split}
		\label{eq:shocktube-initial}
		\left(\rho_\mathrm{L}, p_\mathrm{L}, B^y_\mathrm{L}\right) &= \left(1, 1, 0.5\right),\\
		\left(\rho_\mathrm{R}, p_\mathrm{R}, B^y_\mathrm{R}\right) &= \left(0.125, 0.1, -0.5\right),
	\end{split}
\end{equation}
and all other variables are set to zero.
We evolve the initial conditions until $t = 0.4$, at which point the left-going
rarefaction wave and right-going shock wave have nearly reached the simulation
boundaries.

\begin{figure}[t]
	\centering
	\includegraphics[width=\linewidth]{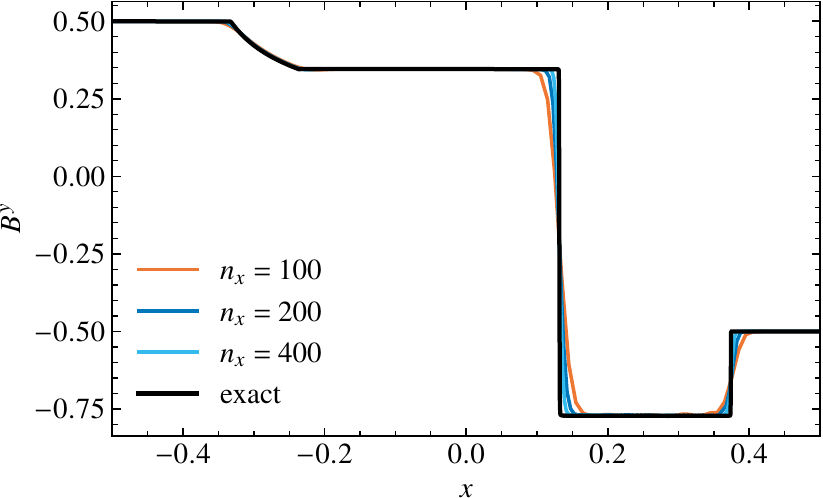}
	\caption {The $y$-component of the magnetic field is shown at $t = 0.4$
	for the shock tube test at three resolutions using $\sigma_0 = \num{e6}$. 
    The highest resolution run, the cyan line with $n_x = 400$, is the
    closest to replicating the reference solution.}
    \label{fig:shocktube-varyres}
\end{figure}

\begin{figure}[t]
	\centering
	\includegraphics[width=\linewidth]{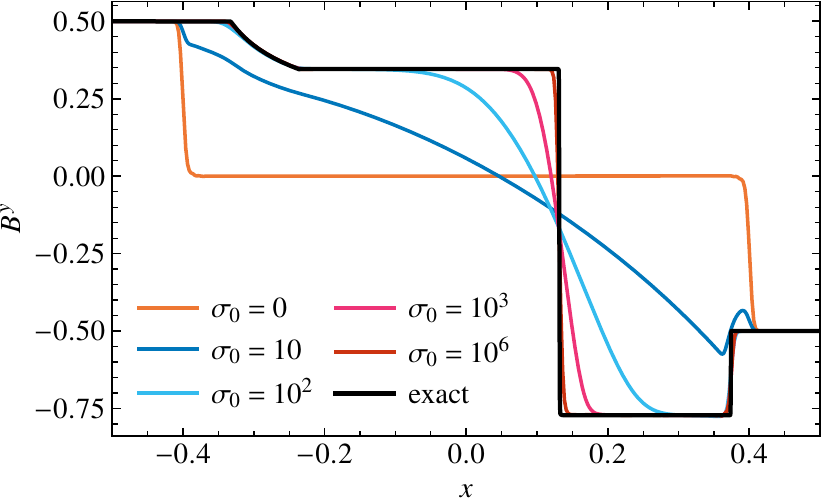}
	\caption{The $y$-component of the magnetic field is shown at $t = 0.4$ for the
	shock tube test using five values of constant uniform conductivity $\sigma_0$.
    The black line shows the reference solution.
	}
	\label{fig:shocktube-varycond}
\end{figure}

\Cref{fig:shocktube-varyres} shows the $y$-component of the magnetic field at
$t=0.4$ for $\sigma_0 = 10^6$ at three resolutions ($n_x = 100, 200, 400$)
compared to a high-resolution ($n_x = 8000$) ``exact'' reference solution using
\bam{}'s ideal MHD module.
We observe that a higher-resolution simulation leads to better capturing of
shocks and discontinuities.
In \cref{fig:shocktube-varycond}, the result at $t = 0.4$ for the fixed
resolution of $n_x = 400$ and for the range of uniform conductivities $\sigma_0
= \{0, 10, 10^2, 10^3, 10^6\}$ is shown, again alongside the high-resolution ideal-MHD
reference solution.
We see that, as expected, the highest conductivity $\sigma_0 = \num{e6}$
most closely approaches the ideal limit, and values $\sigma_0 \lesssim \num{e2}$
give highly dissipative results.
Repeating the test with $\sigma_0 > \num{e6}$ showed no accuracy increase in
reproducing the ideal-MHD reference solution.
With $\sigma_0 = 0$, corresponding to the electrovacuum limit, we see the
initial discontinuity propagating at the speed of light according to the Maxwell
equations in vacuum.
Our results are equivalent to those seen in the literature
(\eg\cite{Dionysopoulou2013,Azizi2025}).

\begin{figure}[t]
    \centering
    \includegraphics[width=\linewidth]{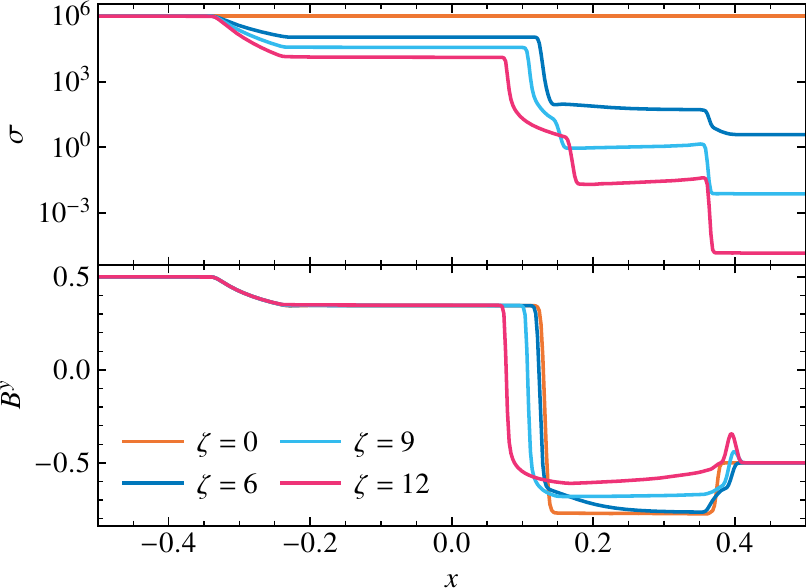}
	\caption{Shock tube tests with the same initial conditions as in
	\cref{fig:shocktube-varyres} and \cref{fig:shocktube-varycond} but with the
	power-law conductivity prescription of \cref{eq:shocktube-powerlaw}.}
	\label{fig:shocktube-powerlaw}
\end{figure}

Since we intend to encode physical information in the conductivity parameter
$\sigma$, which will cover many orders of magnitude across the simulation
grid, it is important to test how the code handles a variable conductivity.
Following \cite{Dionysopoulou2013}, we utilize a profile of the form
\begin{equation}
	\sigma = \sigma_0 \left( \frac{D}{D_0} \right)^\zeta,
	\label{eq:shocktube-powerlaw}
\end{equation}
where \(D_0 = 1\) is a reference value of the conserved rest-mass density \(D\) and
\(\zeta\) is a constant integer exponent.
We again run the shock tube to $t = 0.4$ with the same initial conditions given
in \cref{eq:shocktube-initial} with $n_x = 400$ and $\sigma_0 = \num{e6}$ for
four choices of $\zeta$, namely $\zeta = \{0, 6, 9, 12\}$, and show the results in \cref{fig:shocktube-powerlaw}.
The top panel of \cref{fig:shocktube-powerlaw} shows the value of $\sigma$ as
a function of $x$ at $t = 0.4$ and how it couples with the conserved rest-mass
density.
For large $\zeta$, the simulation tends to the $\sigma = 10^6$ limit on the left (which implies
that the solution for the magnetic field should approach the
ideal-MHD limit in that region), and
the vacuum limit on the right, with the conductivity spanning about 11 orders
of magnitude.
The bottom panel of \cref{fig:shocktube-powerlaw} shows $B^y$ for each value of
$\zeta$, which all show the same rarefaction wave on the left; on the right,
the exact behavior of the shock depends on the conductivity.
For instance, the $\zeta = 12$ case has an additional peak corresponding to the
wavelike behavior of the magnetic field in the vacuum limit.
These shock tube tests demonstrate that our code can handle a large range of
constant and variable conductivities in the presence of one-dimensional shocks.

\subsection{Magnetic monopole test}\label{sec:magmono}
In this test, we simulate an artificially introduced magnetic monopole to
observe the effectiveness and explore the parameters of the divergence cleaning
scheme previously described.
We expect that the process in which the monopole is exponentially damped and
advected to the simulation boundary to depend on the magnetic and electric
divergence cleaning parameters $\kappa_\phi$ and $\kappa_\psi$.
For simplicity, we fix them to take the same value, $\kappa$, for the
remainder of this test.
The damping rate of the auxiliary fields $\phi$ and $\psi$ is proportional to
$\kappa$, so that higher $\kappa$ values lead to faster constraint damping. 
However, haphazardly increasing $\kappa$ yields the risk of introducing
stiffness into the evolution equations for $\phi$ and $\psi$ \cite{Moesta2014},
variables for which an explicit time integration is employed. 
Hence, overly large values of $\kappa$ also have to be avoided.

\begin{figure*}[t]
	\centering
	\includegraphics[width=0.7\linewidth]{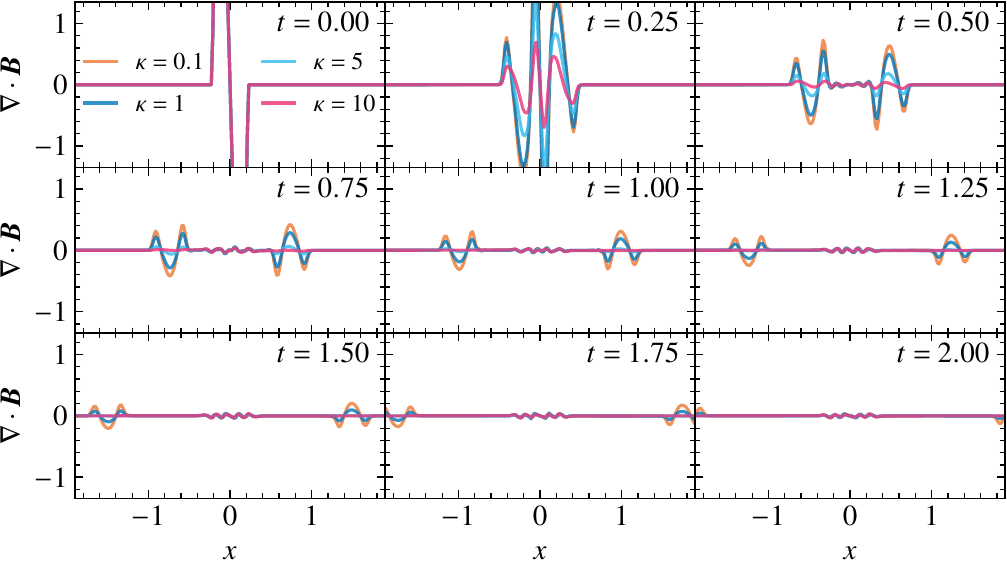}
    \caption{One-dimensional slice along the $x$ axis of the divergence of the
    magnetic field.
    The panels progress forward in time in row-major order.
    The effect of the damped wave equation is clearly visible as the wave
    propagates outwards.
    We find that the small residual in the center of the grid at the end of the
    simulation is further reduced when using a lower-order reconstruction
    scheme, such as LINTVD.}
	\label{fig:magmono-divB}
\end{figure*}

We initialize a Gaussian point monopole with $B^x \neq 0$ in a small spherical
region surrounding the origin following \cite{Moesta2014,Cheong2022}.
Specifically, we set
\begin{equation}
	B^x\left(r\right) = \begin{cases}
		\exp\left(-r^2/R_G^2\right) - \exp\left(-1\right) & \text{if \(r < R_G\)}\\
		0 & \text{if \(r \geq R_G\)}
	\end{cases},
\end{equation}
where \(r^2 = x^2 + y^2 + z^2\) is the spherical radius and \(R_G\) is the
radius of the Gaussian monopole (which is set as 0.2).
On a three-dimensional grid with 200 points in each direction spanning $x,y,z
\in [-2, 2]$, giving a grid spacing of $\Delta x = \Delta y = \Delta z = 0.02$,
we evolve the initial conditions until $t = 2$ for $\kappa = \{0.1, 1, 5, 10\}$.
\Cref{fig:magmono-divB} shows a slice of $\nabla_i B^i$, the divergence of the
magnetic field, along the $x$-axis.
The effect of the damped wave equation obeyed by $\phi$ can be seen in the
evolution of the monopole, which propagates outwards from the origin and
exponentially decays on $1/\kappa$ timescales.
We note that at the end of our simulation, time $t=2.00$, we still have a residual in the central part of the grid, which is consistent with the results presented in \cite{Moesta2014}.
When using the linear total variation diminishing (LINTVD) reconstruction
instead of the higher-order WENOZ reconstruction, the residual is one order of
magnitude smaller.

\begin{figure}[t]
	\centering
	\includegraphics[width=\linewidth]{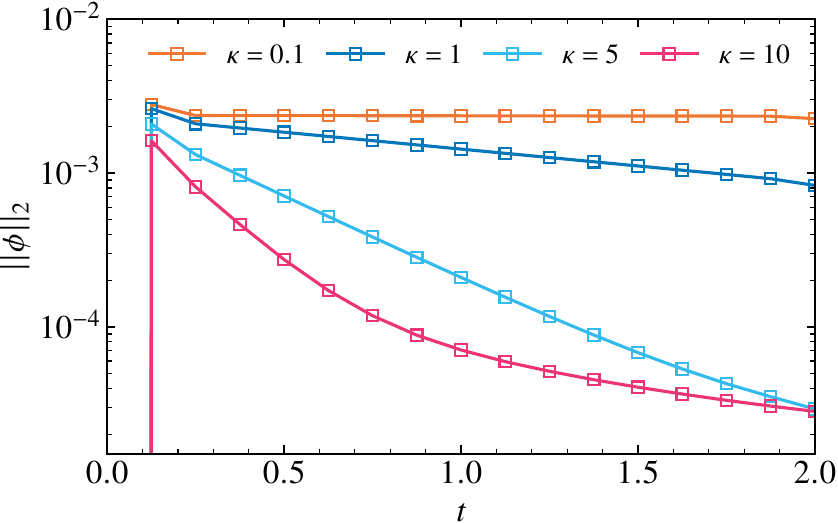}
	\caption
	{\(L^2\) norm of the magnetic divergence cleaning variable \(\phi\) across
	the full three-dimensional domain as a function of time.
	The norm decreases at a faster rate when increasing the value of \(\kappa\).
    Note that continuing to increase the value of $\kappa$ yields a stiffer
    system of equations.}
	\label{fig:magmono-phi}
\end{figure}

The behavior of the divergence control scalar $\phi$ can be seen in
\cref{fig:magmono-phi}, which shows the $L^2$ norm of $\phi$ as computed across
the entire simulation domain.
$\kappa = 5$ and $\kappa = 10$ show similar effectiveness in reducing
$\phi$ as much as possible by $t = 2$, whereas far less damping is exhibited for
$\kappa = 0.1$ and $\kappa = 1$.
It is possibly more effective to choose larger values of $\kappa$, but how
exactly the effectiveness translates from this artificial test problem to real
simulations is unclear.
For symmetry with the ideal GRMHD project in \bam{}, to match other resistive
GRMHD projects using hyperbolic divergence cleaning (\eg\cite{Dionysopoulou2013,
Azizi2025}), we adopt $\kappa_\phi = \kappa_\psi = 1$ for all other simulations
presented in this work.
In the resistive GRMHD simulations of BNS mergers presented in
\cite{Dionysopoulou2015}, the comparatively tiny $\kappa = 0.075$ is used.

\subsection{Circularly polarized Alfv\'en wave}\label{sec:cpalfven}
The propagation of a circularly polarized Alfv\'en wave in one-dimension through a
uniform background magnetic field $B_0$ is another common test in MHD codes.
It is often discussed in the context of ideal MHD (\eg\cite{Del_Zanna2007}), but
it is also possible to perform the test as a probe of a resistive MHD code in
the limit of high conductivity (\eg\cite{Palenzuela2009, Dionysopoulou2013,
Azizi2025}) where one expects to recover the general behavior of ideal MHD.
As we expect the wave to be perfectly advected, we apply periodic boundary conditions to a one-dimensional domain and compare the shape of the wave after one cycle to the initial conditions.

The specific setup is as follows: an Alfv\'en wave with amplitude $\eta_A$ travels
along the $x$-axis, for which the exact solution for the magnetic field can be
written~\cite{Dionysopoulou2013}
\begin{equation}
    	B^i =
        \begin{Bmatrix}
            B_0\\
            B_0\eta_\mathrm{A}\cos\left[k\left(x - v_\mathrm{A} t\right)\right]\\
        	B_0\eta_\mathrm{A}\sin\left[k\left(x - v_\mathrm{A} t\right)\right]
        \end{Bmatrix},
    	\label{eq:alfven-solution}
\end{equation}
where $B_0$ is the strength of the uniform background magnetic field,
$\eta_\mathrm{A}$ is the amplitude of the wave, and \(k = |\bm{k}|\) is the
magnitude of the wavevector.
For simplicity, we choose $v^x = 0$, and the other two components are initially set to
\begin{equation}
	v^y = -v_\mathrm{A} \frac{B^y}{B_0},\qquad
	v^z = -v_\mathrm{A} \frac{B^z}{B_0},
\end{equation}
where the special-relativistic Alfv\'en speed \(v_\mathrm{A}\) is given
by~\cite{Del_Zanna2007}
\begin{equation}
	v_\mathrm{A}^2 = \frac{2B_0^2}{\rho h + B_0^2\left(1 + \eta_\mathrm{A}^2\right)}
		\left[1 + \sqrt{1 - \left(\frac{2\eta_\mathrm{A} B_0^2}
			{\rho h + B_0^2\left(1 + \eta_\mathrm{A}^2\right)}\right)^2}\right]^{-1}.
	\label{eq:alfven-speed}
\end{equation}
One obtains $v_\mathrm{A} = 0.5$ for the specific choices $\rho = p =
\eta_\mathrm{A} = 1$ and $B_0 = 1.1547$.
Furthermore, by choosing a spatial extent $x \in [-0.5, 0.5]$ such that the total
length $L_x = 1$ with $k = 2\pi$, the wave will return to its starting position
at \(t = L_x / v_\mathrm{A} = 2\).

\begin{figure}[bt]
	\centering
	\includegraphics[width=\linewidth]{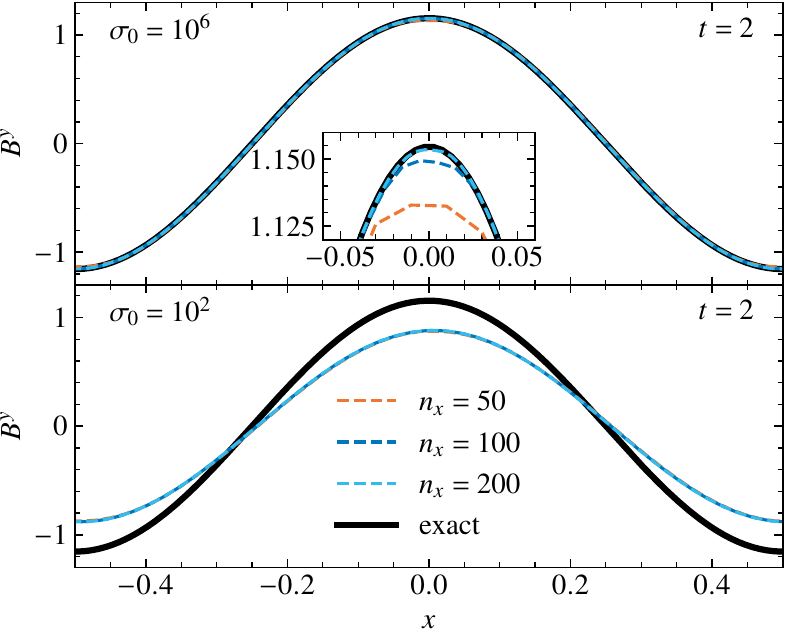}
	\caption
	{Polarized Alfvén wave after one complete
    period \(t = 2\).
    The top panel shows the $\sigma_0 = 10^6$ case, with the zoomed inset
    illustrating the converging solution with increasing resolution. 
    The bottom panel shows the $\sigma_0=10^2$ case, which shows that the
    $\sigma_0=10^2$ case can no longer be considered an accurate approximation
    for ideal MHD.}
	\label{fig:cp-alfven}
\end{figure}

It is important to note that the exact solution \cref{eq:alfven-solution} is
only valid in the ideal-MHD limit; we cannot expect a resistive code to exactly
replicate results from ideal MHD for any finite value of \(\sigma\).
As a close approximation, we perform simulations with a uniform conductivity
$\sigma_0 = \num{e6}$ at three resolutions ($n_x = 50, 100, 200$).
We also run the same simulations with $\sigma_0 = \num{e2}$ for comparison.
The $y$-component of the magnetic field for each resolution after one period is
compared against the exact solution in \cref{fig:cp-alfven}.
The top panel shows that the high conductivity test with $\sigma_0 = \num{e6}$
successfully recovers the ideal MHD exact solution with converging accuracy, and
the bottom plot shows that a lower $\sigma_0$ is far too dissipative to maintain
the expected profile, even after just one period.

\subsection{Self-similar current sheet}\label{sec:currentsheet}
The diffusive behavior observed in the $\sigma_0 = \num{e2}$ circularly
polarized Alfvén wave in the previous section is an expected physical feature of
resistive MHD, and thus something we also would like to test quantitatively.
One can compare the actual diffusion of the magnetic field to theoretical values in
the self-similar current sheet test, first proposed by \cite{Komissarov2007}.

\begin{figure}[t]
	\centering
	\includegraphics[width=\linewidth]{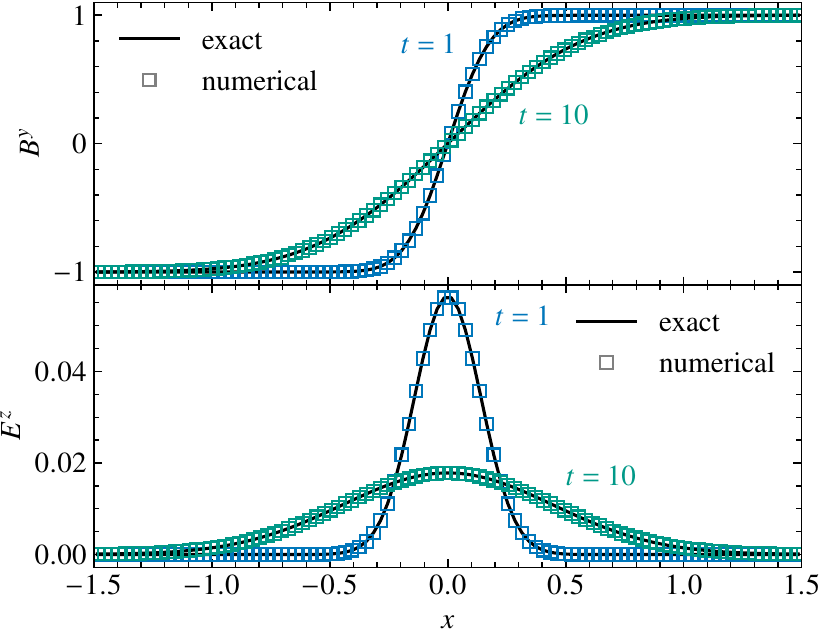}
	\caption
	{The profile of the magnetic field component \(B^y\) (top panel) and the
	implicitly evolved electric field component \(E^z\) (bottom panel) from the
	self-similar current sheet test.
	The result is shown for the initial condition at \(t = 1\) and the evolved
	system at \(t = 10\), using 100 grid points for \(x \in [-1.5, 1.5]\).
    The exact solution is accurately matched.}
	\label{fig:current-sheet}
\end{figure}

The current sheet is composed of a magnetic field in the $y$-direction
that changes sign in a thin layer.
The system is otherwise set up to be in equilibrium, with zero fluid velocity,
uniform constant density ($\rho = 1$) and pressure ($p = 5000$, following
\cite{Cheong2022}), and a moderate conductivity ($\sigma_0 = 100$), which is constant and uniform.
In this state, provided that the fluid pressure dominates over the magnetic
pressure ($\beta_\mathrm{P} = p_\mathrm{gas} / p_\mathrm{mag} \gg 1$), the
electromagnetic field evolves in a self-similar fashion according to the
diffusion equation and the exact solution for $t > 0$ can be written
\cite{Cheong2022}
\begin{equation}
	B^y = \mathrm{erf}\left(\frac{x}{2}\sqrt{\frac{\sigma_0}{t}}\right), \qquad	E^z = \frac{1}{\sqrt{\pi\sigma_0 t}}\exp\left(-\frac{x^2\sigma_0}{4t}\right),
\end{equation}
where $\mathrm{erf}$ is the error function.
The simulation grid contains 200 grid points with $x \in [-1.5, 1.5]$.
The initial data is set according to the equations above; we evolve from $t = 1$
(to avoid dividing by zero) until $t = 10$ and plot $B^y$ and $E^z$ alongside
the exact solutions in \cref{fig:current-sheet}, both of which show excellent
agreement.

\begin{figure}[t]
	\centering
	\includegraphics[width=\linewidth]{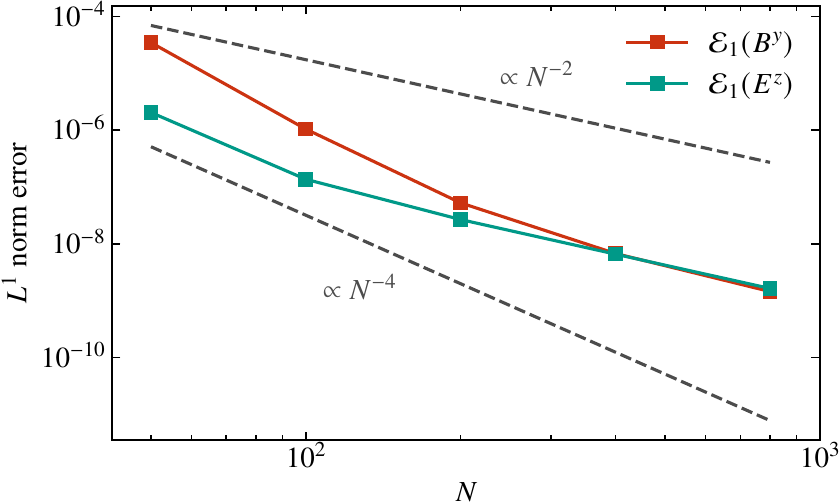}
	\caption
	{Spatial convergence for the self-similar current sheet test.
	The red line indicates the \(L^1\) norm error for the magnetic field component
	\(B^y\) and the teal line for the electric field component \(E^z\).
	The two dashed lines show second- and fourth-order convergence.}
	\label{fig:current-sheet-convergence}
\end{figure}

While it would be possible to check code convergence using the analytic solution, it
is technically only valid for $p \rightarrow \infty$.
Following \cite{Bucciantini2013,Cheong2022}, we avoid using the analytical
solution as a reference and instead use a high-resolution run (\(N = 8000\)) in
its place.
The staggered assignment of coordinates to grid cells in \bam{} means that spatial
interpolation is required to compare identical coordinate points from simulations run at different resolutions.
We thus perform cubic interpolation on the high-resolution reference solutions
\(B^y_\mathrm{ref}\) and \(E^z_\mathrm{ref}\) and measure the \(L^1\) norm error
according to
\begin{equation}
	\label{eq:L1normerror}
	\mathcal{E}_1(Q) = \frac{1}{N} \sum_i ||Q(x_i) - Q_\mathrm{ref}(x_i)||_1,
\end{equation}
where \(Q\) is the quantity of interest, \(Q_\mathrm{ref}\) its value according
to the reference solution, and \(||\cdot||_1\) denotes the \(L^1\) norm.
\Cref{fig:current-sheet-convergence} shows the \(L^1\) norm error for the
\(B^y\) and \(E^z\) as a function of the number of grid points \(N\).
Both quantities clearly exhibit a convergence rate between second order and
fourth order.
At lower resolutions, where spatial error dominates over temporal error, we
achieve approximately fourth-order convergence.
However, as we increase to higher resolutions and spatial inaccuracies become
progressively less important, the convergence rate tends to second-order, which
is what we expect from the second-order IMEX scheme in use.

\subsection{Cylindrical blast wave}\label{sec:cyl-blastwave}
The magnetized blast wave tests are frequently performed in relativistic MHD
codes for their relative simplicity and demanding physical conditions.
Typically performed in either two or three dimensions, these tests consist of an
outwardly propagating cylindrical or spherical explosion in the presence of an
initially uniform magnetic field.
Even though there are no analytic solutions for such problems, they still have
very useful properties.
The oblique shocks that develop as a result of the explosion provide a robust
test of the numerical implementation of shock handling in strongly magnetized
environments.
This test has also been seen in most relativistic resistive MHD implementation
projects
(\eg\cite{Palenzuela2009,Mizuno2013,Dionysopoulou2013,Miranda-Aranguren2018,Mignone2019,Azizi2025}).

We perform the test in the $xy$-plane for $x, y \in [-6, 6]$ for initial data
consisting of three regions.
The inner region is contained within a circle $0 \leq r \leq 0.8$, in which the
density and pressure are highest, namely $\rho = 0.01$ and $p = 1$.
This inner pressurized region is surrounded by the intermediate region, $0.8 < r
< 1$, in which the density and pressure exponentially decrease to match the
surrounding region, where $\rho = p = 0.001$.
We set an initial magnetic field along the $x$-axis with a strength $B_0 =
0.05$. 

\begin{figure}[ht]
    \centering
    \includegraphics[width=\linewidth]{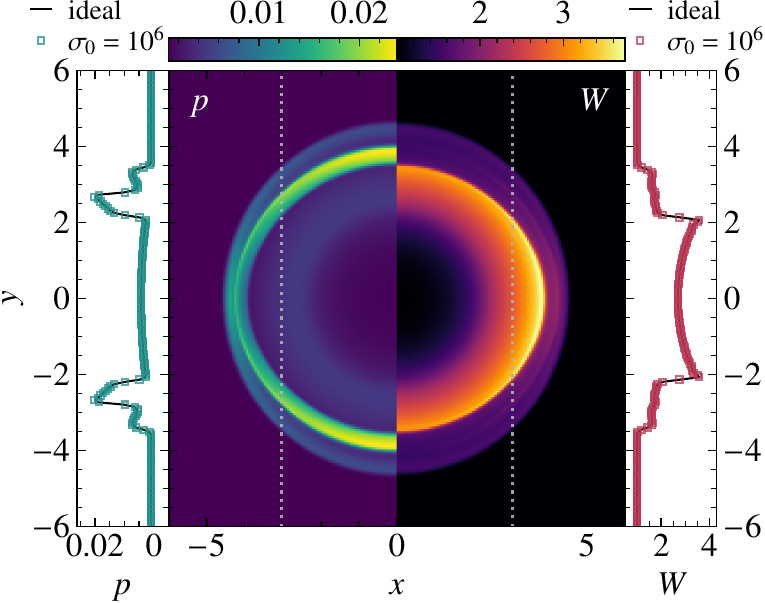}
	\caption{Result from the cylindrical blast wave test shown at \(t = 4\).
	The center panel shows $p$ (left) and $W$ (right) in the $xy$-plane, with
	light gray dotted lines indicating the one-dimensional slices shown
	in the side panels, which are compared to the solution obtained with using
	ideal MHD in \bam{}.}
	\label{fig:cyl-p-W}
\end{figure}
\begin{figure}[ht]
    \centering
    \includegraphics[width=\linewidth]{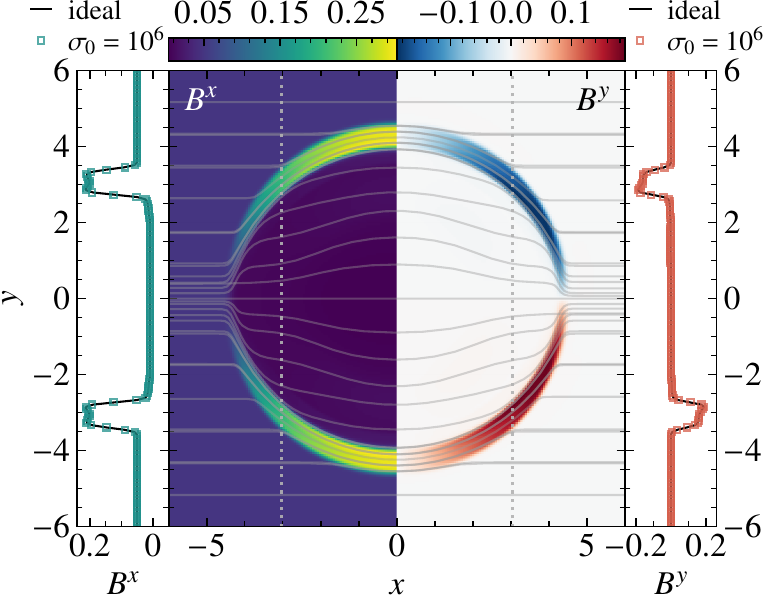}
	\caption{Result from the cylindrical blast wave test shown at \(t = 4\).
	The center panel shows $B^x$ (left) and $B^y$ (right) with field lines
	overplotted in the $xy$-plane, with light gray dotted lines indicating the
	one-dimensional slices shown in the side panels, which are compared
	to the solution obtained using ideal MHD in \bam{}.}
	\label{fig:cyl-Bx-By}
\end{figure}

We use a $200\times200$ grid and a uniform constant conductivity $\sigma_0 =
\num{e6}$ and show the results of the simulation at $t = 4$ in
\cref{fig:cyl-Bx-By,fig:cyl-p-W}.
The center panel of \cref{fig:cyl-p-W} shows that the blast wave produces a
shock that propagates preferentially in the direction parallel to the magnetic
field (\ie along the $x$-axis), as well as a reverse shock inside of which the
motion is isotropic.
The expanding fluid moves preferentially along $x$, developing a higher maximum Lorentz
factor ($W_\mathrm{max} \simeq 3.3$), but there is stronger buildup of fluid
pressure along $y$.
The center panel of \cref{fig:cyl-Bx-By} shows the build up of magnetic
pressure in the $y$-direction as magnetic field lines pile up in the outgoing
shell.
The side panels of \cref{fig:cyl-Bx-By,fig:cyl-p-W} show that the resistive code
very accurately reproduces the ideal MHD result.
We also ran the cylindrical blast wave test with smaller values of $\sigma_0$
and observed only minor differences; however, as expected, the similarity to the
ideal MHD solution is maximized for large values of $\sigma_0$.

\subsection{Spherical blast wave}
The spherical blast wave is a three-dimensional extension of the cylindrical
case and typically one uses identical initial conditions
\cite{Komissarov2007,Mignone2019}, but replaces the cylindrical radius by the
spherical radius $r = \sqrt{x^2 + y^2 + z^2}$.
However, we encountered numerical instability under these conditions for the spherical explosion.
This is a consequence of updating the primitive variables during the implicit
step following the method of \cite{Palenzuela2009}, who report that the
one-dimensional primitive recovery schemes for relativistic resistive MHD codes
are demonstrably unstable for high magnetizations $\sigma_\mathrm{mag} = b^2
/\rho$ and high Lorentz factors \cite{Ripperda2019a}.
We therefore use a reduced magnetic field $B_0 = 0.04$, slightly lower than the
value of $0.05$ used in \cite{Dionysopoulou2013} and 40\% of the value used in
\cite{Mignone2019}.
This is done to reduce the maximum magnetization to ensure stability; reducing
the inner and outer densities achieve the same effect.
The remaining initial conditions and simulation parameters remain the same as in
the cylindrical case, namely $\sigma_0 = \num{e6}$, $\rho_\mathrm{in} =
\num{e-2}$, $\rho_\mathrm{out} = \num{e-3}$.
However, the reconstruction scheme was changed from WENOZ to LINTVD to avoid
artificial oscillations in the charge.
With $B_0 = 0.1$, numerical instability was also encountered in
\cite{Mignone2019} for $\sigma_0 > \num{e2}$ when using hyperbolic
divergence cleaning, but they did not report any stability issues with their
constrained transport variant.

\begin{figure}[t]
	\centering
	\includegraphics[width=0.8\linewidth]{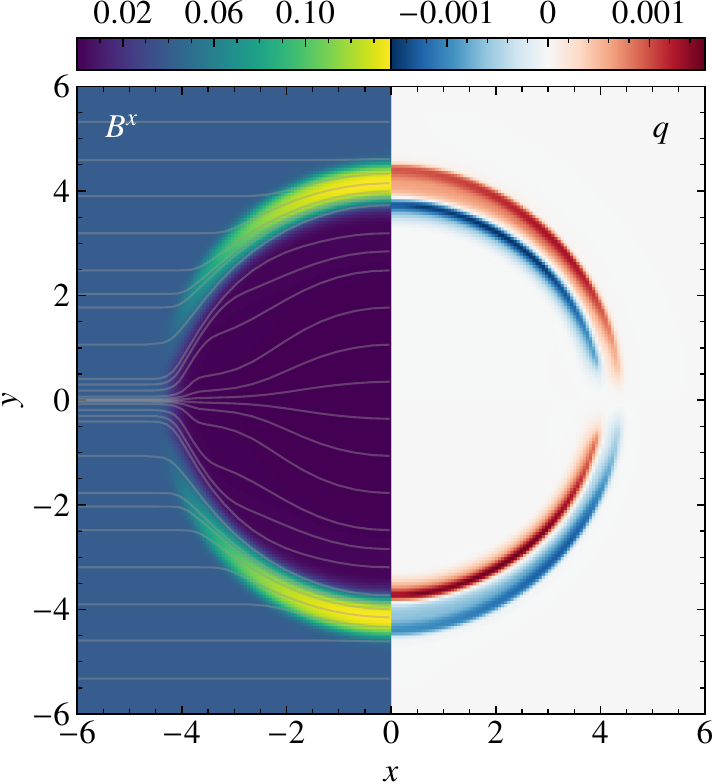}
	\caption
	{Result from the spherical blast wave test.
	The left panel shows $B^x$ and magnetic field lines, and the right panel
	shows the local charge density $q$ generated by the spherical blast
	wave.}
    \label{fig:sph-Bx-q}
\end{figure}

\begin{figure}[ht]
    \centering
    \includegraphics[width=\linewidth]{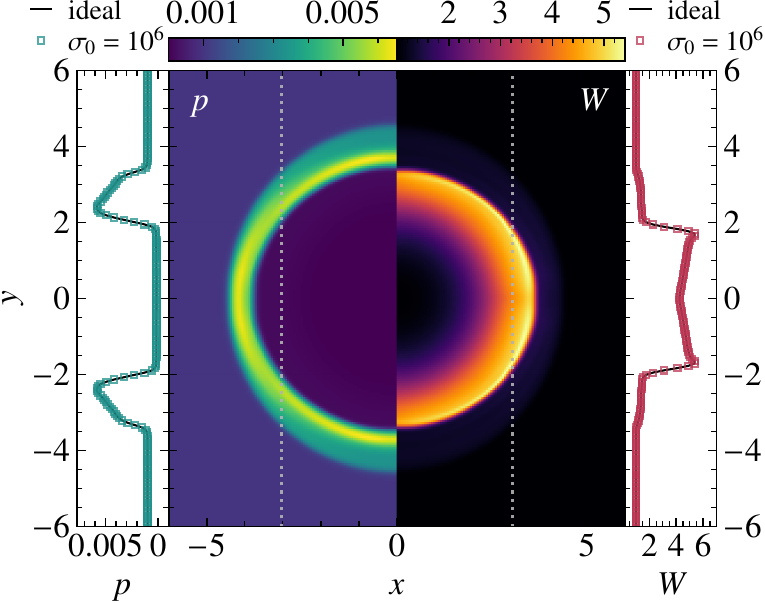}
	\caption
	{Result from the spherical blast wave test.
	The center panel shows $p$ (left) and $W$ (right) in the $xy$-plane, with
	light gray dotted lines indicating the one-dimensional slices shown
	in the side panels, which are compared to the solution obtained with using
	ideal MHD in \bam{}.}
    \label{fig:sph-p-W}
\end{figure}

We run the simulation on a three-dimensional $200\times200\times200$ grid
until $t = 4$.
Despite similar conditions, there are some differences to the planar case.
\Cref{fig:sph-Bx-q} shows $B^x$ and the charge density $q$, which is no longer
identically zero, as $E^z$, the only non-vanishing component of the electric field,
no longer has a null directional derivative.
We observe the same bunching of magnetic field lines in the $y$-direction as in
the cylindrical case.
In \cref{fig:sph-p-W}, we see that the outgoing shock moves faster than in the
cylindrical case ($W_\mathrm{max} \simeq 5.4$) despite the lower-order
reconstruction scheme and lesser magnetization, and the outgoing pressure wave
is more isotropic.
Furthermore, the side panels of \cref{fig:sph-p-W} indicate that the result
here achieves very good agreement with the ideal MHD case, even on the
high--Lorentz factor shock front.

\subsection{Stationary charged vortex}\label{sec:chargedvortex}
This two-dimensional test consists of a rotating flow with uniform density
embedded in a vertical magnetic field.
It is notable for being an exact equilibrium solution of the relativistic
resistive MHD equations, and was first introduced in \cite{Mignone2019}.
This test was also performed in \cite{Ripperda2019a,Cheong2022,Mignone2024}.
The vertical magnetic field and rotating flow give rise to a radial electric field
which also maintains a distribution of electric charge.
In standard cylindrical coordinates \((r, \varphi, z)\), the initial data
can be written \cite{Mignone2019}
\begin{align}
	\label{eq:cv-Er}
	E^r &= \frac{q_0}{2}\frac{r}{r^2 + 1},\\
	\label{eq:cv-q}
	q &= \frac{1}{r}\partial_r\left(rE^r\right)
		= \frac{q_0}{\left(r^2 + 1\right)^2},\\
	B^z &= \frac{\sqrt{\left(r^2 + 1\right)^2 - q_0^2 / 4}}{r^2 + 1},\\
	\label{eq:cv-p}
	p &= -\frac{\rho}{\Gamma_1}
		+ \left[\frac{4r^2 + 4 - q_0^2}
			{\left(r^2 + 1\right)\left(4 - q_0^2\right)}\right]^{\Gamma_1 / 2}
		\left(p_0 + \frac{\rho}{\Gamma_1}\right),\\
	\label{eq:cv-vphi}
	v^\varphi &= -\frac{q_0}{2}
		\frac{r}{\sqrt{\left(r^2 + 1\right)^2 - q_0^2 / 4}},
\end{align}
where \(\Gamma_1 = \Gamma / (\Gamma - 1)\).
Since it is a stationary equilibrium, the initial data are also the exact
solution.

We perform the simulation with a uniform density $\rho = 1$, choosing $p_0 =
0.1$ and $q_0 = 0.7$, on a $200 \times 200$ grid spanning $x,y\in[-10, 10]$
until $t = 5$.
The equilibrium condition does not depend on the conductivity parameter, despite
being an exact solution of relativistic resistive MHD, so one is free to choose
$\sigma_0$~\cite{Mignone2019}.
We therefore perform the charged vortex test for $\sigma_0 \in
[\num{e-1}, \num{e7}]$ (uniform base 10--logarithmic spacing) to examine the
stability of high conductivity (and thus highly stiff) simulations.

\begin{figure}[t]
	\centering
	\includegraphics[width=0.8\linewidth]{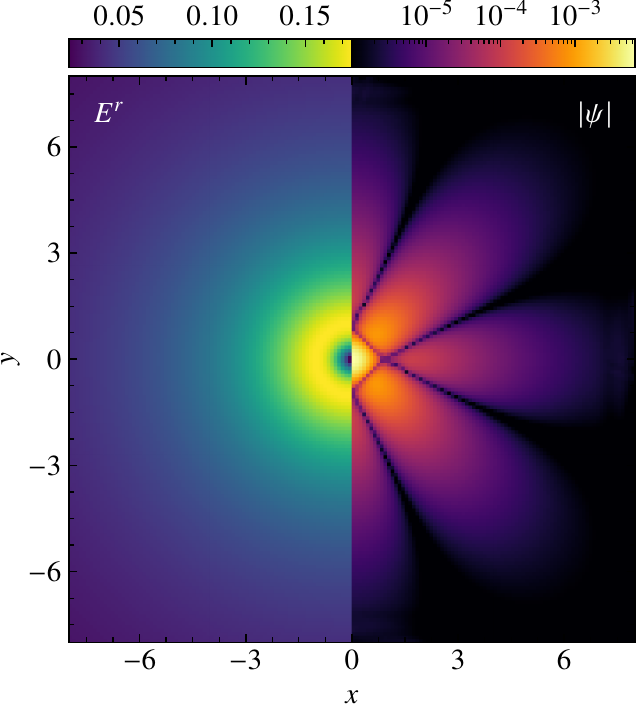}
	\caption
	{Simulation results for the stationary charged vortex.
	The left side shows the radial component of the electric field \(E^r\)
	\cref{eq:cv-Er} and the right side shows the magnitude of the divergence
	control variable \(\psi\).
	Non-zero values of \(\psi\) indicate violations of the constraint \(D_iE^i =
	q\).
	The charge density \(q\) is shown in \cref{fig:charged-vortex-q-vphi}. }
	\label{fig:charged-vortex-Er-psi}
\end{figure}
\begin{figure}[ht]
	\centering
	\includegraphics[width=0.8\linewidth]{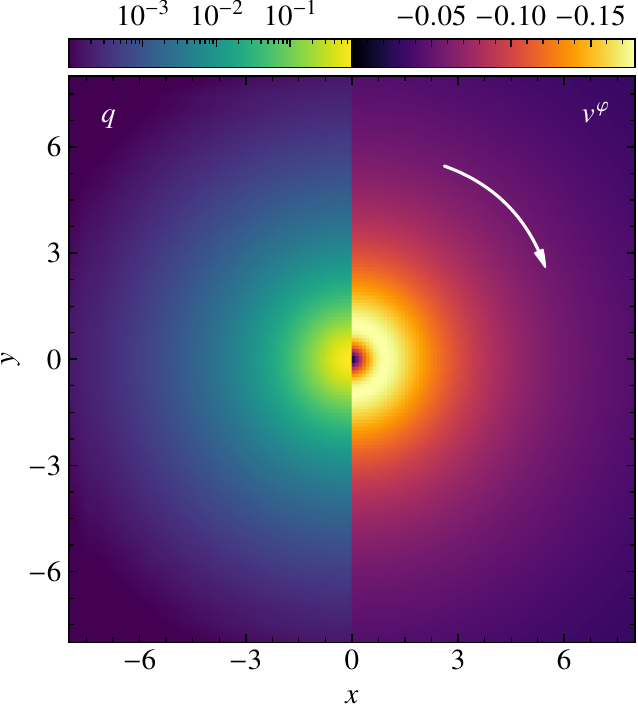}
	\caption
	{Simulation results for the stationary charged vortex.
	The left panel shows the electric charge density \(q\) \cref{eq:cv-q} and
	the right panel shows the azimuthal velocity \(v^\varphi\) \cref{eq:cv-vphi},
	with an indicator arrow illustrating the direction of the flow.}
	\label{fig:charged-vortex-q-vphi}
\end{figure}

The radial electric field $E^r$ and the magnitude of the divergence control
scalar $\psi$ at $t = 5$ are shown in \cref{fig:charged-vortex-Er-psi}.
Non-zero values of $\psi$ correspond to violations of the constraint $D_iE^i =
q$; however, the relative magnitude of the violations are small.
Similarly, \cref{fig:charged-vortex-q-vphi} shows the charge density $q$ and
the azimuthal velocity $v^\varphi$, with an arrow on the latter plot to indicate
the direction of motion.
\begin{figure}[t]
	\centering
	\includegraphics[width=\linewidth]{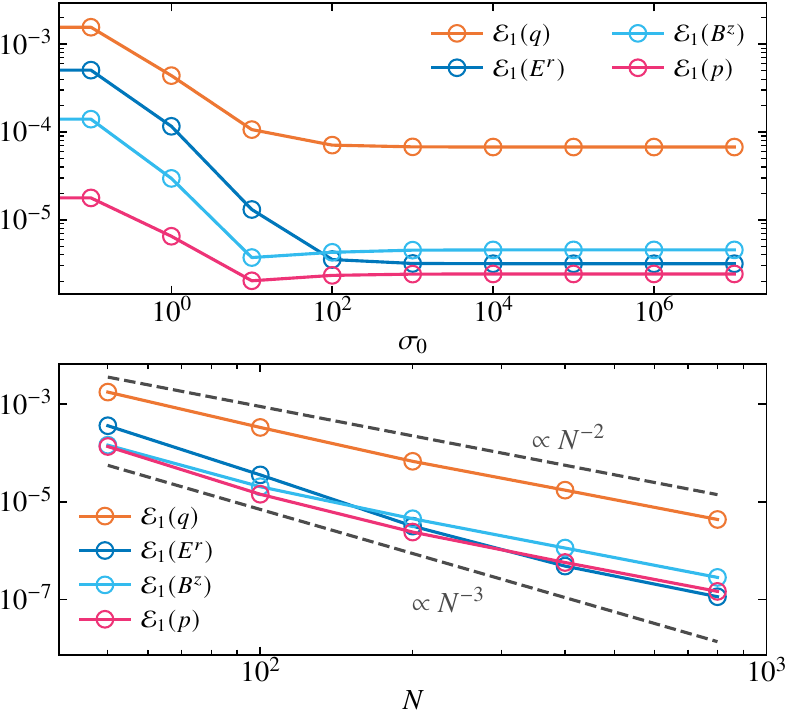}
	\caption[Convergence plots for the charged vortex test]%
	{Convergence plots for the stationary charged vortex test.
	The \(L^1\) norm error, computed at \(t = 5\), is shown for the quantities
	\(q\), \(E^r\), \(B^z\), and \(p\), as a function of conductivity \(\sigma_0\)
	in the top panel and as a function of grid size (\(N \times N\) points) in the
	bottom panel (with $\sigma_0 = \num{e3}$).
	All four variables shown exhibit a spatial convergence rate between second and third order.}
	\label{fig:charged-vortex-convergence}
\end{figure}
The $L^1$ norm error as defined in \cref{eq:L1normerror} for the quantities $q$, $E^r$, $B^z$, and $p$ at $t = 5$
are shown in the bottom panel of \cref{fig:charged-vortex-convergence}, all of
which achieve between second- and third-order convergence, demonstrating that
our code successfully maintains charge density in an equilibrium configuration.
The top panel of \cref{fig:charged-vortex-convergence} shows that as the
conductivity increases, we see a decrease in the \(L^1\) norm error of \(q\),
\(E^r\), \(B^z\), and \(p\) until the threshold value of \(\sigma_0 \simeq \num{e2}\).
Although the equilibrium state is independent of $\sigma_0$, it still affects 
the diffusivity of the plasma. 
These results suggest that around this threshold, the physical diffusivity
encoded by the conductivity parameter becomes small compared to the inherent
numerical diffusivity present in the scheme at this resolution.
Consequently, we see nearly identical results for $\sigma_0 \geq \num{e2}$.

\subsection{Two-dimensional relativistic rotor}
The relativistic rotor or magnetic rotor is a non-equilibrium two-dimensional
test for relativistic MHD codes,
first proposed in~\cite{Del_Zanna2003} for the ideal MHD case.
The resistive version, first presented in \cite{Dumbser2009}, is ubiquitous in
special- and general-relativistic MHD literature
(\eg\cite{Bucciantini2013, Miranda-Aranguren2018, Shibata2021, Nakamura2023,
Mignone2024}).
Although no analytical solution exists, it is nonetheless a useful test to
perform, as the outcome depends strongly on the conductivity.

\begin{figure*}[htb]
	\centering
	\includegraphics[width=0.95\textwidth]{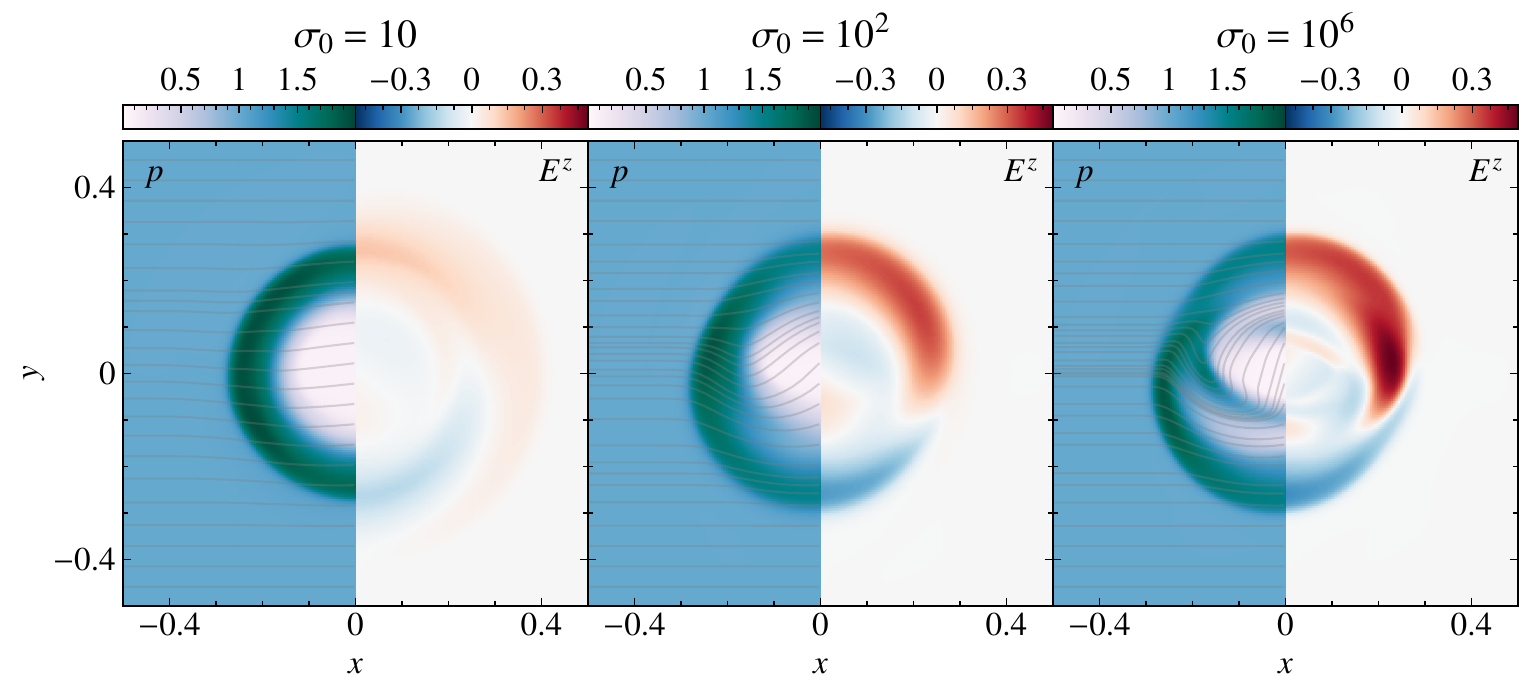}
	\caption
	{Two-dimensional plots of the pressure \(p\) (left) and $z$-component of
	the electric field \(E^z\) (right) for the relativistic rotor.
	The three columns correspond to different values of $\sigma_0$, as indicated
	by top text.
    Magnetic field lines are plotted on top of the pressure and show how the
    coupling of the fluid and the electromagnetic field becomes stronger as we
    approach the ideal-MHD limit.}
	\label{fig:magrot-p-Ez}
\end{figure*}
\begin{figure*}[htb]
	\centering
	\includegraphics[width=0.95\textwidth]{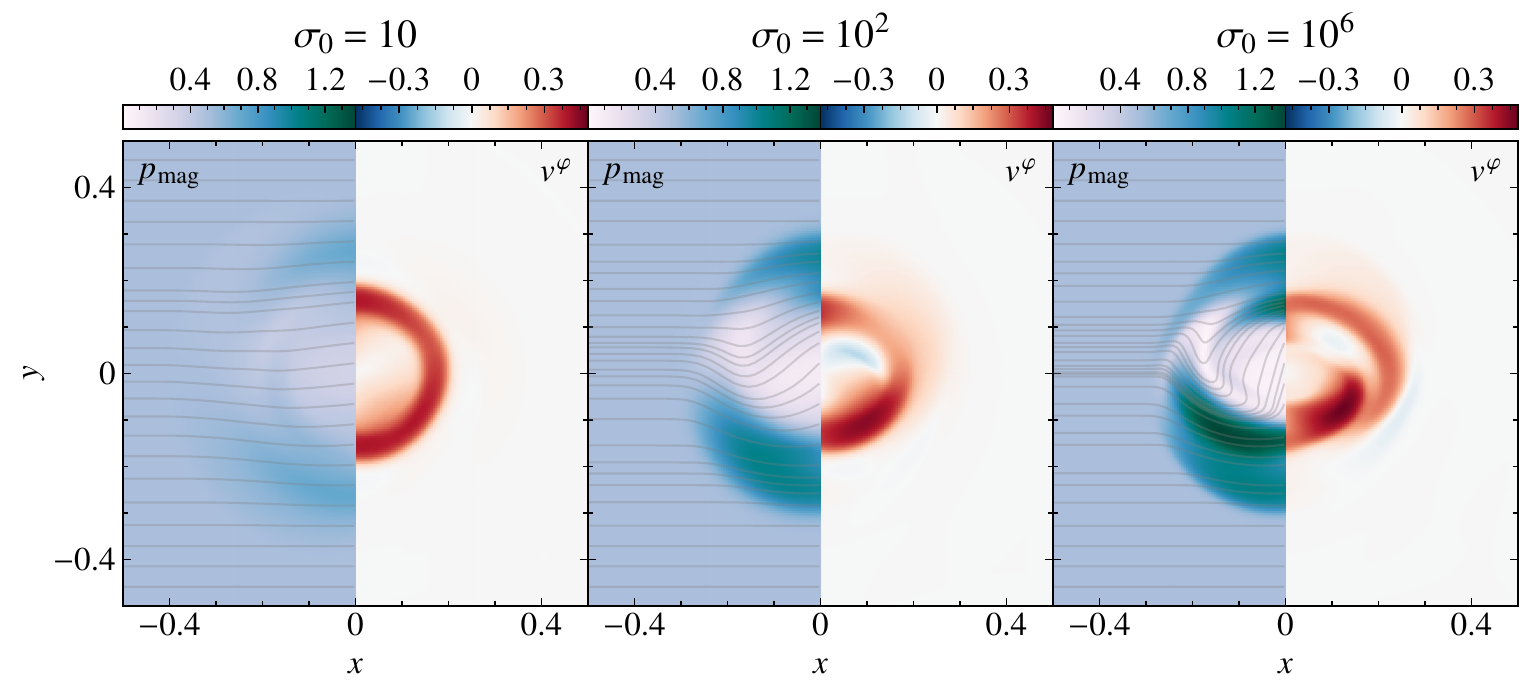}
	\caption
	{Two-dimensional plots of the magnetic pressure \(p_\mathrm{mag}\) (left) and
	the azimuthal velocity \(v^\varphi\) (right) for the relativistic rotor.
	The three columns correspond to different values of $\sigma_0$, as indicated
	by top text.
	Magnetic field lines are plotted on top of the magnetic pressure.}
	\label{fig:magrot-pmag-vphi}
\end{figure*}

On a square, two-dimensional grid with $x, y \in [-0.5, 0.5]$, we initialize a
fluid clump with $\rho = 10$ within a circular region $r = \sqrt{x^2 + y^2} <
0.1$.
The clump is rapidly rotating with a uniform angular velocity $\Omega
= 8.5$ (\ie $v^x = -\Omega y$ and $v^y = \Omega x$) inside a static ($v^i = 0$)
and uniform background region with $\rho = 1$ and $p = 1$.
A constant uniform magnetic field $B^i = (1, 0, 0)$ fills the domain and
surrounds the rotor.
The initial electric field is computed according to the ideal-MHD definition
$E^i = -\epsilon^{ijk}v_j B_k$.
As the rotor spins, the magnetic field progressively wraps itself around it,
which launches torsional Alfv\'en waves into the surrounding ambient fluid
\cite{Mignone2024}.
At the final time ($t = 0.3$), the initially circular clump has been deformed
into its characteristic oval shape with a severity that depends on the conductivity.

We simulate the relativistic rotor on a $200\times200$ grid for three increasing
values of uniform conductivity until $t = 0.3$.
To perform this test in the high-conductivity \(\sigma_0 = 10^6\) limit without
any crashes, the reconstruction scheme was changed to LINTVD from WENOZ.
For consistency, this was done for all values of $\sigma_0$.
The three cases correspond to the fully resistive regime ($\sigma_0 = 10$), the
quasi-ideal regime ($\sigma_0 = 10^3$), and (approximately) the ideal limit
($\sigma_0 = 10^6$).
\Cref{fig:magrot-p-Ez} shows the fluid pressure and \(z\)-component of the
electric field for the three selected values of \(\sigma_0\) at \(t = 0.3\).
Also, in \cref{fig:magrot-pmag-vphi} we depict the magnetic pressure $p_{\rm
mag}$ and the azimuthal velocity $v^\varphi$ at $t=0.3$ with the three previously
selected values of $\sigma_0$.
With decreasing conductivity, the angular momentum of the rotor as well as the
torsional Alfv\'en waves that cause the deformation are more strongly
dissipated, which in the end gives a much more circular result (see the leftmost
column of \cref{fig:magrot-p-Ez} or \cref{fig:magrot-pmag-vphi}).
For high conductivity ($\sigma_0 = 10^6$), we expect the dynamics of
ideal MHD to apply, such that the fluid and the electromagnetic fields evolve tightly coupled
and thus the torsional Alfv\'en waves are highly deforming.
This can be clearly seen in the right columns of
\cref{fig:magrot-p-Ez,fig:magrot-pmag-vphi}.

\subsection{Kelvin--Helmholtz instability}\label{sec:khi}
Ref.~\cite{Rasio1999} suggested the Kelvin--Helmholtz instability (KHI) could
significantly amplify the magnetic field during and after binary neutron star
mergers and \cite{Price2006} showed that amplification by multiple orders of
magnitude is possible. 
High spatial resolution is required in order to simulate this process fully,
since the growth rate of the KHI is proportional to the wave number ($k \propto
1/L$) of the turbulent mode.
Although local simulations of the shear layer at merger can resolve this
\cite{Obergaulinger2010}, global simulations struggle to reach
the required resolutions, but are still able to resolve some amplification via
the turbulent motion of the KHI \cite{Kiuchi2014,Kiuchi2015}.
The KHI is common as a test for ideal MHD codes (\eg\cite{Mignone2009,
Beckwith2011, Kiuchi2022, Neuweiler2024}) and has also been performed for
resistive MHD codes \cite{Mizuno2013,Mignone2019,Mignone2024}.
For our version of the test, we use the same setup as
\cite{Kiuchi2022} and \cite{Neuweiler2024}.
We consider a two-dimensional rectangular simulation grid with $x \in [-0.5,
0.5]$ and $y \in [-1, 1]$.
In order to investigate the effect of resolution on vortex formation and
subsequent turbulence and magnetic-field amplification, we perform
both ``low-resolution'' simulations ($N_x = 100$ and $N_y = 200$) and
``high-resolution'' simulations ($N_x = 200$ and $N_y = 400$).
For each resolution, we test with three uniform conductivities $\sigma_0 =
\{10^3, 10^4, 10^6\}$.
The grid is initialized with uniform pressure and density of $p = 20$ and $\rho
= 1$ and a velocity in the $x$-direction with $\tanh$-shaped profile of the form
\begin{equation}
	v^x = -v_\mathrm{sh} \tanh\left(\frac{y}{a}\right),
\end{equation}
where $v_\mathrm{sh}$ is the shear velocity and $a$ measures the thickness of
the shear layer.
The velocity in the $y$-direction is initialized to disturb the shear layer:
\begin{equation}
	v^y = A_0 v_\mathrm{sh}\sin\left(2\pi x\right) \exp\left(-100y^2\right),
\end{equation}
where $A_0$ is a parameter that controls the strength of the perturbation.
There is no initial velocity in the $z$-direction.
Following \cite{Kiuchi2022,Neuweiler2024} we choose $v_\mathrm{sh} = 0.25$, $a =
0.02$, and $A_0 = \num{e-4}$.
The initial magnetic field is given by
\begin{equation}
	\left(B^x, B^y, B^z\right) = \left(\sqrt{2 p \sigma_\mathrm{pol}}, 0,
	0\right),
\end{equation}
where we choose the poloidal magnetization to be $\sigma_\mathrm{pol} = 0.01$;
the magnetic field is uniform and parallel to the velocity field in the lower
half of the domain.
For each case we evolve the initial conditions until $t = 20$.

\begin{figure*}[ht]
	\centering
	\includegraphics[width=0.92\textwidth]{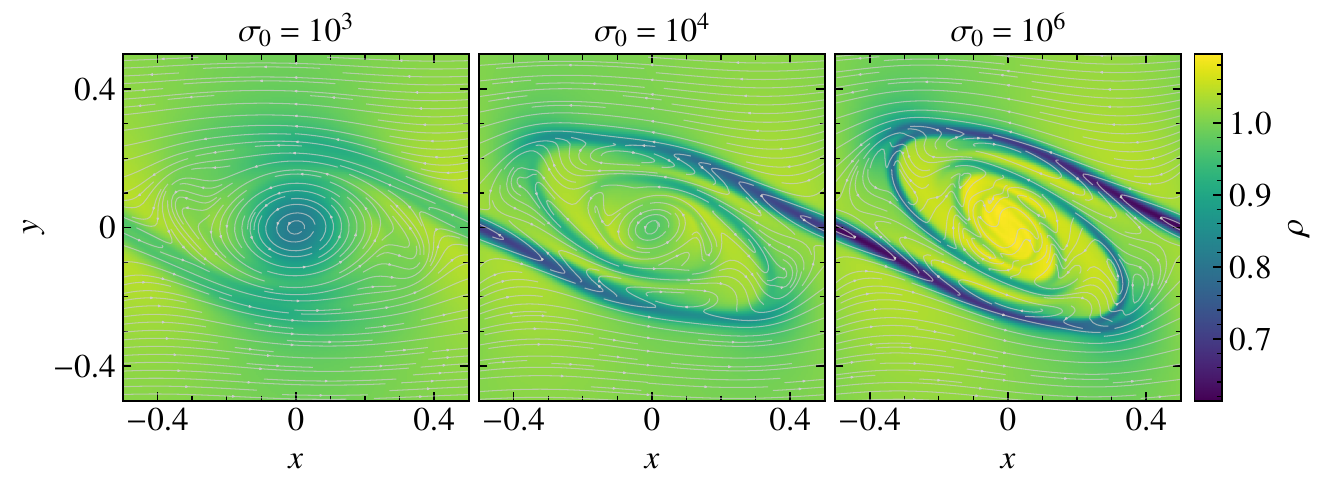}
	\caption[Vortex formation in the Kelvin--Helmholtz instability (KHI) test]%
	{Density plots showing vortex formation via the Kelvin--Helmholtz
	instability for three values of uniform conductivity $\sigma_0$ at $t = 15$.
	The left panel with $\sigma_0 = \num{e3}$ represents the intermediate
	conductivity regime; the center panel with $\sigma_0 = \num{e4}$ shows
	the approach to the ideal MHD limit; the right panel with $\sigma_0 =
	\num{e6}$ accurately approximates the ideal MHD regime.
	The white lines indicate the velocity field and show the bulk motion as well
	as the turbulent motion inside the vortex.}
	\label{fig:khi-vortex}
\end{figure*}

\begin{figure}[th]
	\centering
	\includegraphics[width=\linewidth]{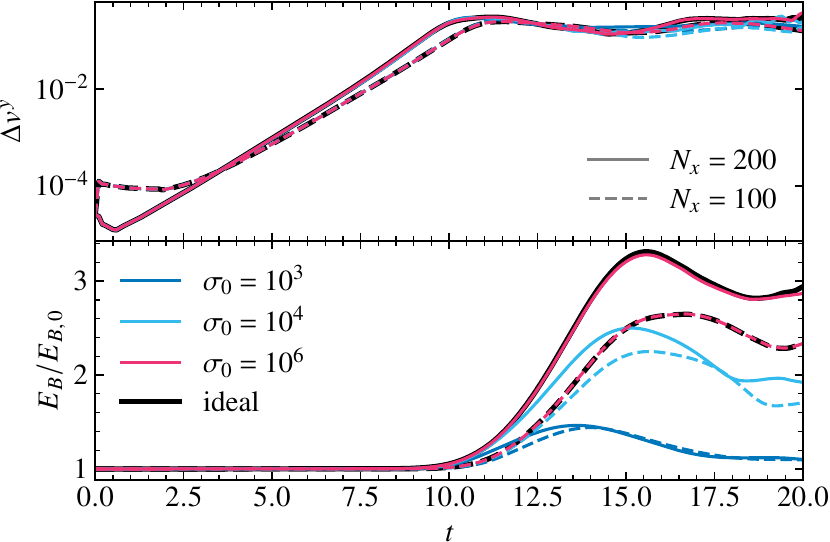}
	\caption{
    The top panel shows the perturbed velocity difference $\Delta v^y =
    0.5(v^y_\mathrm{max} - v^y_\mathrm{min})$ and the bottom panel shows the
    magnetic field energy $E_B$ as a function of time for the Kelvin--Helmholtz
    instability test.
	$\Delta y$ has no dependence on the conductivity parameter $\sigma_0$ and only
	depends on the resolution. $E_B$, however, strongly depends on $\sigma_0$ and
	the ability to resolve tight turbulent vortexes.
    Setting $\sigma_0 = \num{e6}$ produces results nearly identical to the ideal
    MHD module in \bam{}.
	}
	\label{fig:khi-growth}
\end{figure}

\Cref{fig:khi-vortex} shows $\rho$ and the velocity field lines for the
high-resolutions for each value of $\sigma_0$.
The ability to form a vortex via the KHI evidently depends on the strength of
the coupling between the fluid and the electromagnetic fields via the
conductivity, similar to the observations in the previous section.
The lowest conductivity run with $\sigma_0 = 10^3$ is
barely able to resolve the vortex at all, and further tests with $\sigma_0 <
\num{e3}$ show that as $\sigma_0$ approaches zero, the tightness of the winding
in the vortex decreases as well.
Note that these simulations were performed with the local Lax--Friedrichs (LLF)
Riemann solver.
The vortex structure seen in \cref{fig:khi-vortex} is different than the ones
seen in \citep{Kiuchi2022,Neuweiler2024}, which both use a less diffusive
Riemann solver than LLF, such as HLLE \cite{Harten1983,Einfeldt1988} or HLLD
\cite{Mignone2009}.
Our results for $\sigma_0 = 10^6$ and for the ideal MHD case, both using LLF,
closely resemble the HLLE results in \cite{Bucciantini2006,Kiuchi2022}.
Nearly identical results were obtained for $\sigma_0 > 10^6$, indicating that
this value for the conductivity is a good choice to represent the ideal MHD
limit.

In \cref{fig:khi-growth}, the evolution of the perturbed velocity difference
\begin{equation}
	\Delta v^y = \frac{1}{2}\left(v^y_\mathrm{max} - v^y_\mathrm{min}\right)
\end{equation}
as well as the amplification of magnetic field energy $E_B$ for both resolutions
and all values of $\sigma_0$ are shown.
For comparison, results from the ideal MHD module in \bam{} are shown by black
lines.
The perturbed velocity difference grows exponentially in all simulations at a
resolution-dependent rate before reaching non-linear saturation, which occurs at
$t \simeq 11.5$ for the lower resolution and at $t \simeq 11.0$ for the higher
resolution.
The growth rate and time of saturation is independent of $\sigma_0$.
This is not the case for the growth of the magnetic field, which begins only
after the onset of nonlinear saturation, as seen in the bottom panel of
\cref{fig:khi-growth}.
The amplification of the magnetic field depends on the conductivity parameter,
which is consistent with amplification process of stretching and folding field
lines, as they are most strongly coupled to the fluid for high $\sigma_0$.
The $\sigma_0 = 10^6$ run matches the ideal case very closely; the
$\sigma_0 = 10^4$ run shows less efficient amplification and the $\sigma_0 = 10^3$
run, which does not fully resolve the vortex, shows initial growth in magnetic field
energy before decaying again due to Ohmic diffusion, which occurs on a
timescales inversely proportional to the resistivity (and thus proportional to
the conductivity).

\subsection{Isolated magnetized neutron star with poloidal fields}\label{sec:tov-pol}
In order to further verify our implementation, we now consider a test in which
the spacetime is evolved in full general relativity.
A common and simple choice is a spherical star whose fluid initial
data is computed as a solution to the Tolman--Oppenheimer--Volkoff (TOV)
equations with corresponding spacetime according to the Einstein field
equations.

Neutron stars are thought to possess dipolar magnetic fields extending far
outside the star~\cite{Goldreich1969}.
In the magnetosphere, the electromagnetic field dynamically dominates the
plasma, and one can treat this region in the force-free 
approximation~\cite{Uchida1997,Komissarov2002,Carrasco2018}.
This limiting case corresponds to the low-inertia limit of relativistic
magnetohydrodynamics ($\rho \rightarrow 0$, $\sigma \rightarrow \infty$).
In contrast, the electrovacuum treatment ($\rho \rightarrow 0$, $\sigma
\rightarrow 0$) is common among similar resistive GRMHD works
(\eg\cite{Dionysopoulou2013,Dionysopoulou2015,Cheong2022,Franceschetti2025,Azizi2025})
but it is less realistic description of the rarefied plasma in the
magnetosphere~\cite{Goldreich1969,Blandford1977} and usually chosen only for
simplicity.
Both cases are seen in the literature (see,
\eg\cite{Lehner2012,Etienne2017,Parfrey2017,Carrasco2018}
for force-free approaches), but some argue that neither of these limiting
approximations are sufficient to explain the morphology and spectra of
high-energy emission from pulsars \cite{Li2012}.

In general, describing the electrical conductivity of a neutron star is far from
trivial (\eg\cite{Uzdensky2011,Pons2019}), but is generally highest at the
center of the star and much lower at the surface and outside.
In order to capture the ideal MHD limit inside the highly-conductive star and
the electrovacuum limit%
\footnote{Note that it is possible to effectively model the magnetosphere
exterior using force-free electrodynamics within our existing framework via a
careful introduction of a phenomenological current density incorporating both
isotropic and anisotropic effects~\cite{Palenzuela2013}.}
outside, with a smooth transition between the
two marked by a sharp decline through the star's ``edge'', following
\cite{Dionysopoulou2013} we adopt the density-tracking conductivity profile
\begin{equation}
	\sigma = \sigma_0 \max\left(1 - \frac{D_\mathrm{thr}}{D}, 0\right)^2,
	\label{eq:sigma-trackD}
\end{equation}
where $D_\mathrm{thr} = f_\mathrm{thr}\rho_\mathrm{atm}$ (see \cref{sec:atm}).
This conductivity profile in combination with our resistive GRMHD formulation
automatically enforces the correct boundary conditions at the surface, without
the need for matching separate interior and exterior fields
(\eg\cite{Baumgarte2003,Lehner2012,Paschalidis2013}).
Since we simulate non-rotating stars, a magnetosphere supported by charges will
not develop, and the electrovacuum exterior is an acceptable approximation.

For these simulations we employ a piecewise-polytropic fit of the
SLy EOS~\cite{Douchin:2001sv} for the star, following the approach of
\cite{Read:2008iy}.
A star with a central density of $\rho_c = \num{1.28e-3} =
\qty{7.88e14}{g~cm^{-3}}$ thus has a gravitational mass of \qty{1.22}{\msun} and
an isotropic radius of \qty{9.58}{km}.
For the evolution, we extend the zero-temperature EOS by a thermal pressure
$P_\mathrm{th} = (\Gamma_\mathrm{th} - 1)\rho
\epsilon_\mathrm{th}$~\cite{Bauswein:2010dn}, with $\Gamma_\mathrm{th} =
1.75\epsilon_\mathrm{th}$ as the thermal part of the specific internal energy.
A poloidal magnetic field confined to the star's interior is superimposed at the
start of the simulation, leading to a small constraint violation at $t=0$, which
is subsequently damped through the Z4c evolution system~\cite{Weyhausen:2011cg}.
To ensure the initial magnetic field is divergence-free, we initialize it from
a toroidal vector potential expressed in cylindrical coordinates as
\begin{equation}
    A_\varphi = r^2 \max\left[A_b\left(p - p_\mathrm{cut}\right), 0\right]^2,
    \label{eq:A-pol}
\end{equation}
where $A_b$ controls the strength of the field and $p_\mathrm{cut}$ is the limiting
pressure.
We choose $A_b = 1$, corresponding to an initial maximum field strength of
$\sim$\qty{2e12}{\gauss}, and $p_\mathrm{cut}$ to be 4\% of the initial central
pressure. 

We perform the simulations in three spatial dimensions with bitant symmetry
across the $z = 0$ plane.
The grid consists of four fixed refinement levels.
To study the effect of grid spacing, the simulations are run at three different
resolutions that each cover the same physical volume.
The lowest resolution consists of 120 points on each level corresponding to a
coarsest and finest grid spacing of \qty{2.36}{km} and \qty{295}{m}, the medium
resolution of 160 points with grid spacings of \qty{1.77}{km} and \qty{221}{m},
and the highest resolution of 192 points with grid spacings of \qty{1.48}{km}
and \qty{185}{m}.
In addition to the multiple resolutions, we also perform simulations for
multiple values the conductivity parameter $\sigma_0$ in
\cref{eq:sigma-trackD}; namely, we use $\sigma_0 = \{\num{e2}, \num{2e2},
\num{e3}, \num{e6}\}$ to roughly cover the parameter space from the
unrealistically diffusive to an approximation of ideal GRMHD.
Each star is evolved until $t = \qty{2440}{\msun} \simeq \qty{12}{ms}$ and a
Courant factor of 0.2 is applied.
For the HRSC treatment we use a WENOZ reconstruction method with a LINTVD
fallback in low-density regions (see Sec.~II~D~2 in \cite{Neuweiler2024}),
and the LLF Riemann solver as described in \cref{sec:riemann}.

\begin{figure}[t]
    \centering
    \includegraphics[width=\linewidth]{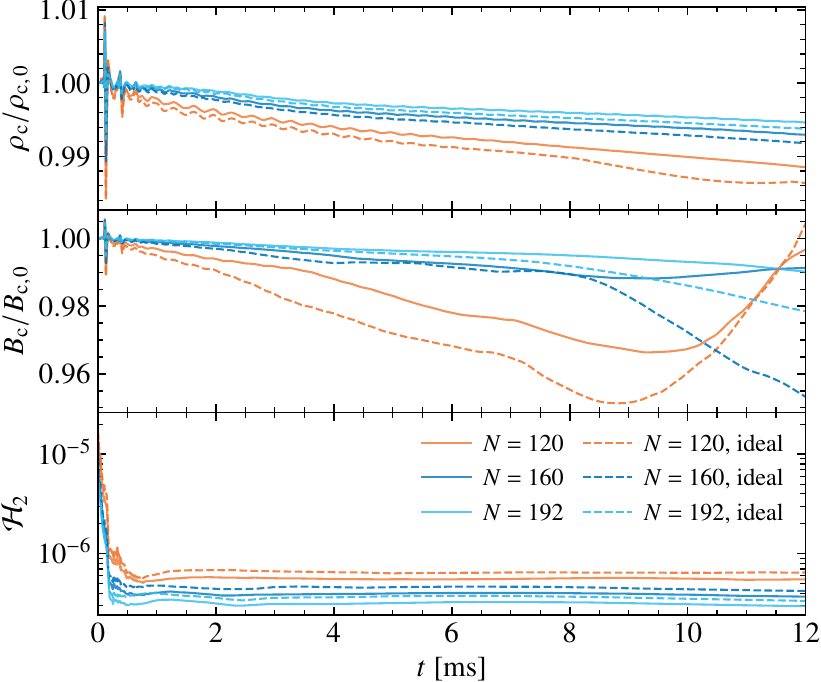}
    \caption{
    Evolution of the TOV star simulation for varying resolutions.
    The different colors indicate the three resolutions used; the solid lines
    show the simulations with $\sigma_0 = \num{e6}$ and the dashed lines show
    the ideal GRMHD runs.
    Displayed is the evolution of the central rest-mass density
    $\rho_\mathrm{c}$ (top panel), the central magnetic field strength
    $B_\mathrm{c}$ (middle panel), and the norm of the Hamiltonian constraint
    $\mathcal{H}_2$ (bottom panel).
    }
    \label{fig:tov-res}
\end{figure}

\begin{figure}[th]
    \centering
    \includegraphics[width=\linewidth]{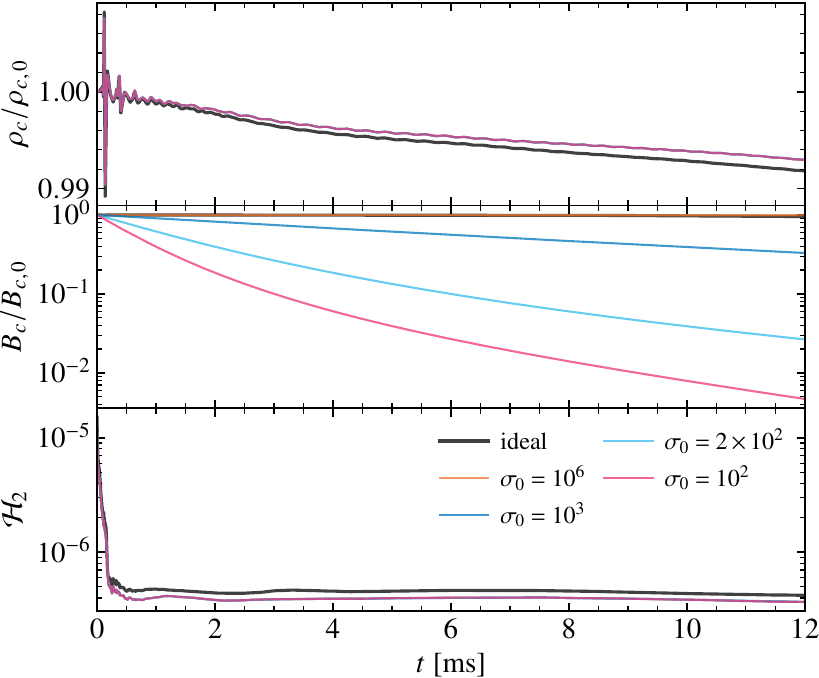}
    \caption{
    Evolution of the TOV star simulation for varying conductivities.
    The different colors indicate the value for $\sigma_0$; the black line
    shows the simulation run with the ideal GRMHD.
    Displayed is the evolution of the central rest-mass density
    $\rho_\mathrm{c}$ (top panel), the central magnetic field strength
    $B_\mathrm{c}$ (middle panel), and the norm of the Hamiltonian constraint
    $\mathcal{H}_2$ (bottom panel).
    Different values of $\sigma_0$ have negligible effects on $\rho_\mathrm{c}$
    and $\mathcal{H}_2$, but $B_\mathrm{c}$ decays according to the Ohmic
    diffusion timescale $t_\mathrm{diff} \propto \sigma_0$.
    }
    \label{fig:tov-sigma}
\end{figure}

\Cref{fig:tov-res} shows the evolution of the isolated neutron stars over the
course of our simulations as we vary the resolution.
We compare, at the three resolutions, the change in the central rest-mass
density $\rho_\mathrm{c}$, the central magnetic field strength $B_\mathrm{c}$,
and the norm of the Hamiltonian constraint $\mathcal{H}_2$ using the resistive
module with $\sigma_0 = \num{e6}$ as well as the ideal GRMHD module.
The central densities (top panel) are observed to decay monotonically with small
fluctuations as the star radially oscillates as a result of truncation error,
with a slight systematic different between the resistive and ideal modules. The
amplitude of these fluctuations depends on resolution.
The behavior of Hamiltonian constraint violations (bottom panel) is similar
across the board, rapidly damped down to an essentially constant level that
decreases as resolution increases.
There is again a slight difference between the resistive and ideal cases.

The central magnetic field (middle panel) also experiences a decaying trend and
is somewhat coupled to the central density.
The field strength may appear to slightly decay due to Ohmic diffusion, but for
$\sigma_0 = \num{e6}$, this effect is comparatively small
(\cf\cref{fig:tov-sigma}); furthermore, since approximately equal decay rates
are observed for both the resistive and ideal cases, we may attribute this
mainly to numerical diffusion.
This is further evidenced by decreasing rates of decay for the three increasing
resolutions.
The central magnetic field experiences changes beyond the expected Ohmic
dissipation at later times which depend on resolution, most likely as a result
of the atmosphere.
As the outer layers of the star expands and the boundary between the edge of the
star and the atmosphere region becomes more extended, the inner layers are
slightly compressed towards the center.
With $\sigma_0 = \num{e6}$ we replicate the fluid-frozen magnetic field lines of
ideal MHD, which leads to the compression of field lines and increase in the
central magnetic field that is not reflected in the total electromagnetic
energy.
The time at which this occurs is resolution dependent.
Similar results were also observed by \cite{Dionysopoulou2013, Azizi2025}.
A note must be made regarding our highly diffusive Riemann solver (LLF) and the
maximally diffusive eigenvalues, which permit, especially at lower resolutions,
the inner structure of the magnetic field to change unpredictably after $t
\sim 7\text{--}9$~\unit{ms} as the fluid in the star undergoes simultaneous
expansion and compression at different radial distance.
Although the resistive module approximates the ideal one, they are not identical
in implementation (as we shall again see in \cref{sec:bns}).
Differences in how exactly the field reconfiguration affects the simulation,
given the implementation differences and presence of a (small) physical
resistivity for the resistive case, may explain the variations between the two
versions of the simulations, which is most apparent for $N = 160$ (\eg the field
strength may be increased by reconnection).

In \cref{fig:tov-sigma}, the results of five simulations with $N = 160$ but
varying $\sigma_0$ are shown, along with the ideal GRMHD result.
The theoretically expected timescale of Ohmic diffusion is
\mbox{$t_\mathrm{diff} = 4 \pi \lambda_B^2\sigma / c^2$} \cite{Palenzuela2009,
Harutyunyan2018}, where $\lambda_B$ is the characteristic lengthscale of
magnetic field variations. 
Since the simulations in resistive GRMHD are identical apart from $\sigma_0$,
the spatial variations are the same, and so the relevant diffusion timescale is
directly proportional to the conductivity.
The negligible change in $B_\mathrm{c}$ for the $\sigma_0 = \num{e6}$ and ideal
cases in \cref{fig:tov-sigma} show that the rate of numerical diffusion is
relatively small, and that the decay observed for the other values of $\sigma_0$
is indeed physically driven Ohmic diffusion rather than a numerical effect.
As an example, consider in the center panel of \cref{fig:tov-sigma} the
$\sigma_0 = \num{e2}$ (magenta line) and $\sigma_0 = \num{2e2}$ (light blue
line) simulations.
The magnetic field strength in the $\sigma_0 = \num{e2}$ case drops by an order
of magnitude by $t \simeq \qty{3}{ms}$.
When $\sigma_0$ is doubled, also doubling the timescale of Ohmic diffusion, a
decay by one order of magnitude from the initial value is achieved by the
expected time $t \simeq \qty{6}{ms}$.

\section{Binary neutron star simulations}\label{sec:bns}
\subsection{Initial configurations} 
Finally, we perform BNS simulations of two different initial configurations.
The same piecewise-polytropic fit of the SLy EOS described \cref{sec:tov-pol} is
employed.
The initial data are constructed with the pseudo-spectral code
\textsc{sgrid}~\cite{Tichy2009,Tichy2012,Dietrich:2015pxa,Tichy2019}.
The main difference between the configurations is that one has higher masses and
forms a black hole shortly after the merger, while the other forms a
hypermassive neutron star (HMNS) remnant.
We label the ``low-mass'' simulations ``LM'' and the ``high-mass'' simulations
``HM''.
For each configuration, we consider three different values for the conductivity
parameter $\sigma_0$,
namely $10^2$, $10^4$ and $10^6$, and the ideal case in which $\sigma
\rightarrow \infty$ (here, the ideal GRMHD module of \bam{} is used). 
The simulations have a short inspiral phase of nearly two orbits and the initial
separation is the same for all setups.

A summary of the different configurations can be seen in \cref{tab:bns_sim}
together with the respective resolutions. 
We evolve the HM simulations with $\sigma_0 = 10^6$ using different resolutions
in order to study its influence on the results. 
The resolutions on the finest refinement level range from \qty{218}{m} for the
lowest-resolution simulations (R1) to \qty{145}{m} for the highest (R4).
These correspond to 128 and 192 points in each dimension on the moving levels.
For all of the following BNS simulations, we use in total seven refinement
levels with the finest four levels being moving boxes. 
We again apply a Courant factor of 0.2.
Akin to the isolated neutron star simulations of \cref{sec:tov-pol}, the
conductivity is computed according to the density-tracking profile shown in
\cref{eq:sigma-trackD}.
An initial poloidal magnetic field is added for each star again following
\cref{eq:A-pol}.
However, we increase the initial strength of the magnetic field to
\qty{2.6e15}{\gauss} by choosing $A_b = 1000$.
In contrast to the single neutron star simulations, for the binaries we use
$f_\mathrm{atm} = \num{e-11}$ and $f_\mathrm{thr} = 1.1$ as atmosphere parameters.
We again use the WENOZ reconstruction method with a LINTVD fallback as
introduced in \cref{sec:tov-pol}.

In the following sections, we present some features of our simulations, keeping
a more detailed discussion of the influence of magnetic fields on the matter
outflow and associated electromagnetic counterparts for future studies.

\begingroup
\renewcommand{\arraystretch}{1.3}
\begin{table}
\centering
\caption{Key parameters for the BNS simulations performed.
From left to right, the columns represent the simulation name, the gravitational
masses of the stars (in solar masses), the maximum conductivity value (in code
units), whether the remnant collapses to a black hole, the number of points per
dimension of the moving box that contains the star, and the resolution of this
finest moving box (in meters). 
The naming scheme denotes ``low-mass'' (LM) and "high-mass" (HM) simulations
with a superscript to indicate the conductivity parameter and a subscript for
the resolution.
All simulations were performed with a piecewise-polytropic fit of the SLy EOS.}
\begin{tabular}{ l c c c c c c }
\hline \hline
Name & $M_1$ [\unit{\msun}] & $M_2$ [\unit{\msun}] & $\sigma_0$ & BH? & $N$ &
$\Delta x$ [m]\\%
\hline
LM$_{\rm R1}^{\sigma2}$ & 1.300 & 1.300 & $10^2$ & \ding{55} & 128 & 218 \\
LM$_{\rm R1}^{\sigma4}$ & 1.300 & 1.300 & $10^4$ & \ding{55} & 128 & 218 \\
LM$_{\rm R1}^{\sigma6}$ & 1.300 & 1.300 & $10^6$ & \ding{55} & 128 & 218 \\
LM$_{\rm R1}^{\rm id}$  & 1.300 & 1.300 & ``$\infty$'' & \ding{55} & 128 & 218 \\
HM$_{\rm R1}^{\sigma2}$ & 1.375 & 1.375 & $10^2$ & \ding{51} & 128 & 218 \\
HM$_{\rm R1}^{\sigma4}$ & 1.375 & 1.375 & $10^4$ & \ding{51} & 128 & 218 \\
HM$_{\rm R1}^{\sigma6}$ & 1.375 & 1.375 & $10^6$ & \ding{51} & 128 & 218 \\
HM$_{\rm R1}^{\rm id}$  & 1.375 & 1.375 & ``$\infty$'' & \ding{51} & 128 & 218 \\
HM$_{\rm R2}^{\sigma6}$ & 1.375 & 1.375 & $10^6$ & \ding{51} & 140 & 199 \\
HM$_{\rm R3}^{\sigma6}$ & 1.375 & 1.375 & $10^6$ & \ding{51} & 160 & 174 \\
HM$_{\rm R4}^{\sigma6}$ & 1.375 & 1.375 & $10^6$ & \ding{51} & 192 & 145 \\
\hline\hline
\end{tabular}
\label{tab:bns_sim}
\end{table}
\endgroup

\subsection{Monitoring quantities}
\begin{figure}
    \centering
    \includegraphics[width=\linewidth]{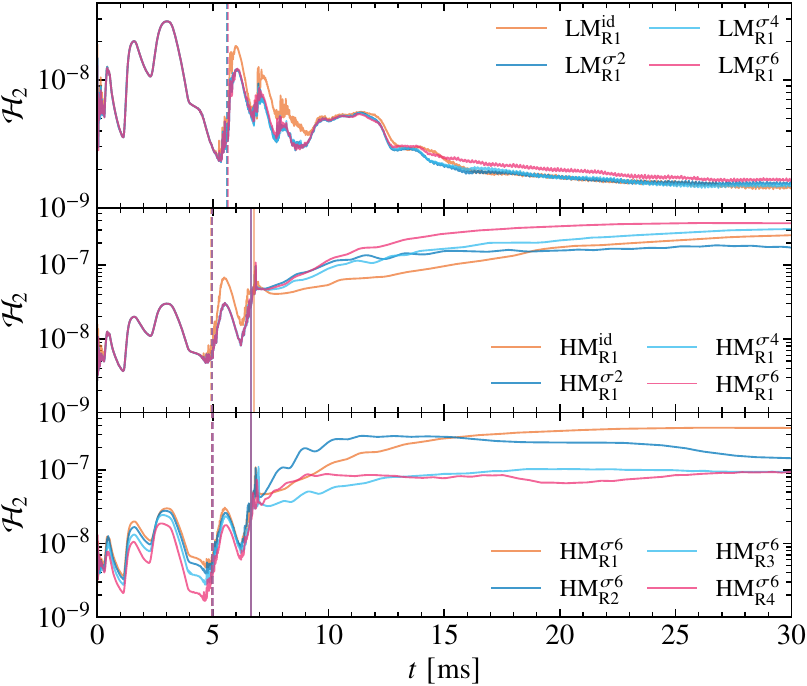}
    \caption{Hamiltonian constraint values for the runs described in
    \cref{tab:bns_sim}.
    In each panel, the dashed vertical lines and the solid
    vertical lines show the merger times and BH formation times, respectively.}
    \label{fig:hamiltonian}
\end{figure}

We first analyze the behavior of the Hamiltonian constraint during the evolution.
In the upper panel of \cref{fig:hamiltonian} we show the values of the
Hamiltonian constraint for the LM runs.
After a quick inspiral phase, the stars merge at $t \sim \qty{5}{ms}$.
The violations of the Hamiltonian constraint remain small at $\sim$\num{e-9}
during the whole simulation for all conductivities.
In the central panel, we clearly see that at the formation of the BH
($\sim$\qty{7}{ms}) the value of the Hamiltonian constraint increases to
$\sim$\num{e-7}, which is typically the case in these situations.
After the BH formation, the Hamiltonian constraint remains above
\num{e-7} which is not a common behavior when compared to simulations without
magnetic fields.
A closer look into two-dimensional plots shows that these higher values come
from inside the black hole horizon and do not affect the evolution.
Finally, in the lower panel, we show that for increasing resolutions the effect
presented in the central panel is reduced.

\subsection{Evolution of electric and magnetic fields}
\begin{figure*}[t]
    \centering
    \includegraphics[width=\linewidth]{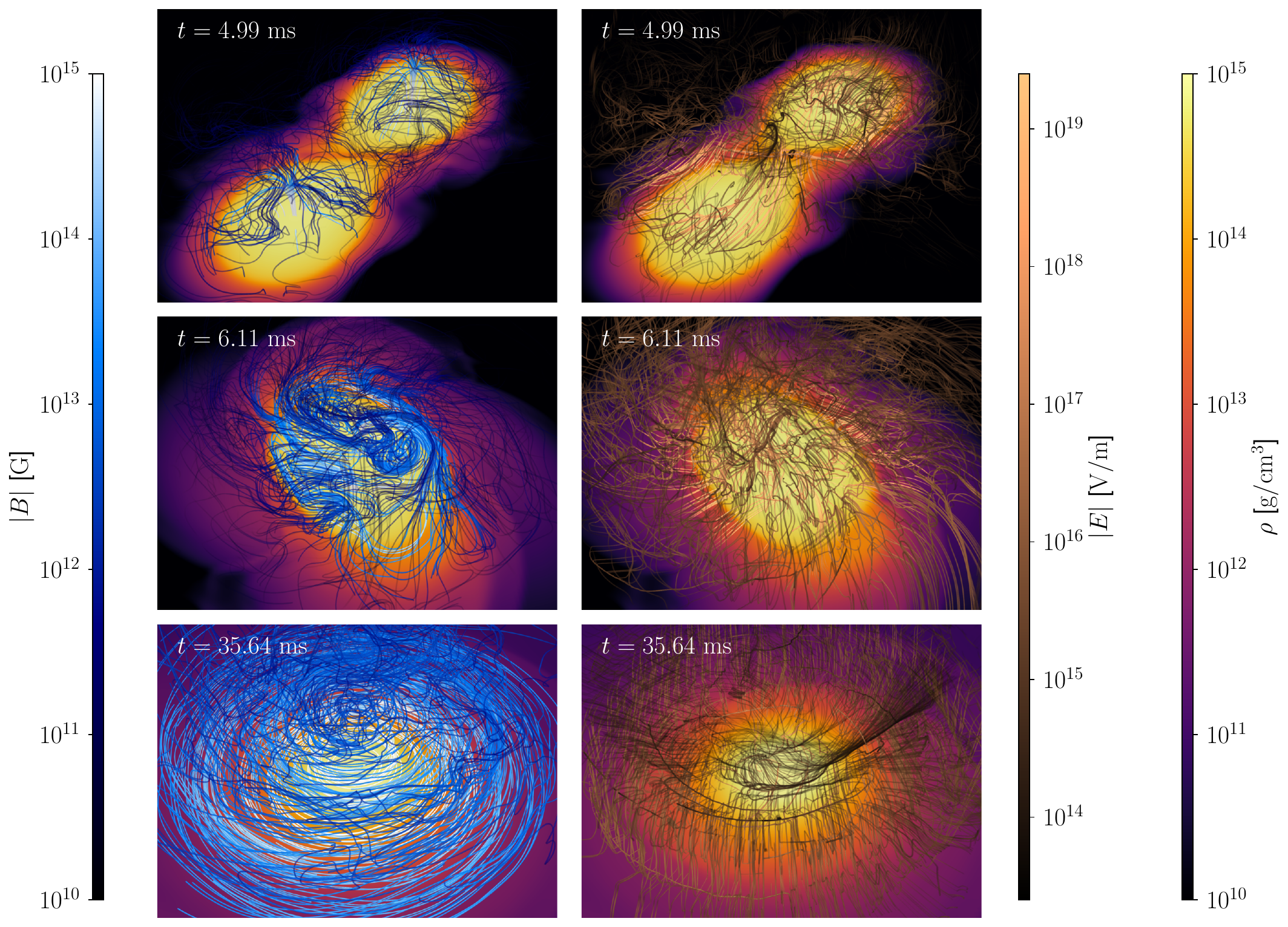}
    \caption{Three-dimensional snapshots of the LM$_\mathrm{R1}^{\sigma6}$
    simulation.
    With a cut at $z=0$, the rest-mass density is shown in the lower half,
    whereas magnetic field lines (left panels) and electric field lines (right
    panels) are displayed in the upper half, respectively.}
    \label{fig:3DBNS-LM}
\end{figure*}

\begin{figure*}[t]
    \centering
    \includegraphics[width=\linewidth]{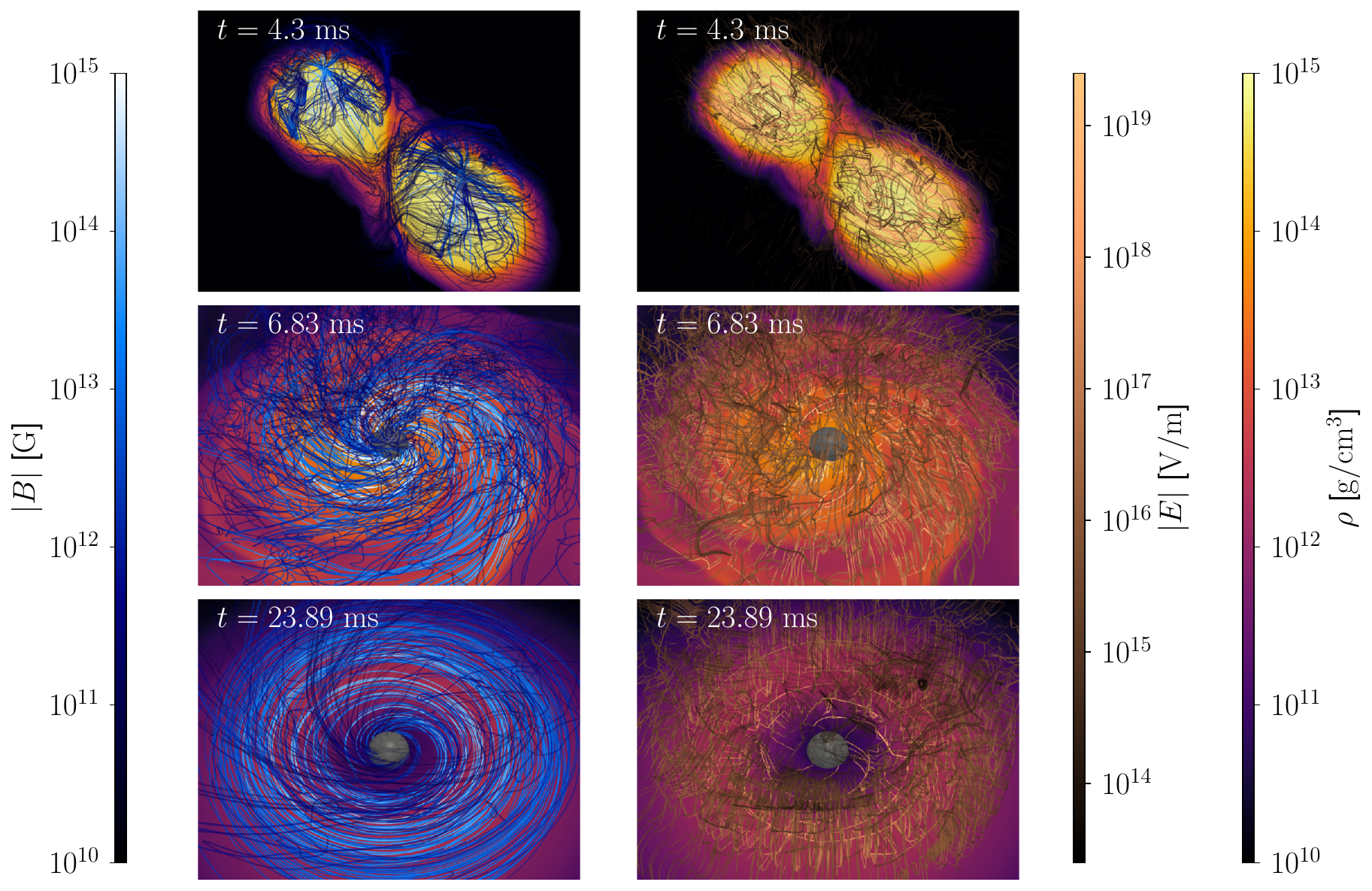}
    \caption{Three-dimensional snapshots of the HM$_\mathrm{R4}^{\sigma6}$
    simulation.
    With a cut at $z=0$, the rest-mass density is shown in the lower half,
    whereas magnetic field lines (left panels) and electric field lines (right
    panels) are displayed in the upper half, respectively.
    Additionally, the apparent horizon of the formed BH is illustrated as a gray
    sphere.}
    \label{fig:3DBNS-HM}
\end{figure*}

For a general overview on the evolution and topology of the electric and
magnetic fields, we provide three-dimensional snapshots of the LM$_{\rm
R1}^{\sigma6}$ and the HM$_{\rm R4}^{\sigma6}$ simulations in
\cref{fig:3DBNS-LM} and \cref{fig:3DBNS-HM}, respectively. 
For both systems, the first snapshots illustrate the fields at merger,
still featuring the initialized poloidal magnetic field structure.
The electric field forms a torus around the center, where the field follows
lines orthogonal to the magnetic field and the stars motion as expected for high
conductivities approaching the ideal GRMHD approximation.
The other panels of \cref{fig:3DBNS-LM} present the topologies during the merger
and post-merger phase with a HMNS remnant.
The turbulence during the collision twists and stretches the magnetic field,
leading to a chaotic structure.
In the ongoing evolution, the remnant's rotation winds the magnetic field and
builds up a mainly toroidal configuration, whereas the electric field forms
spiral structures around the magnetized disk torus.
In contrast, \cref{fig:3DBNS-HM} shows the electric and magnetic field
structures at collapse and in the disk around the remnant BH for the
HM$_\mathrm{R4}^{\sigma6}$ simulation.
Again, we obtain in the disk a toroidal magnetic field with a spiraling electric
field around the disk, while some magnetic field lines penetrate the BH apparent
horizon.

\begin{figure*}[th]
    \centering
    \includegraphics[width=\linewidth]{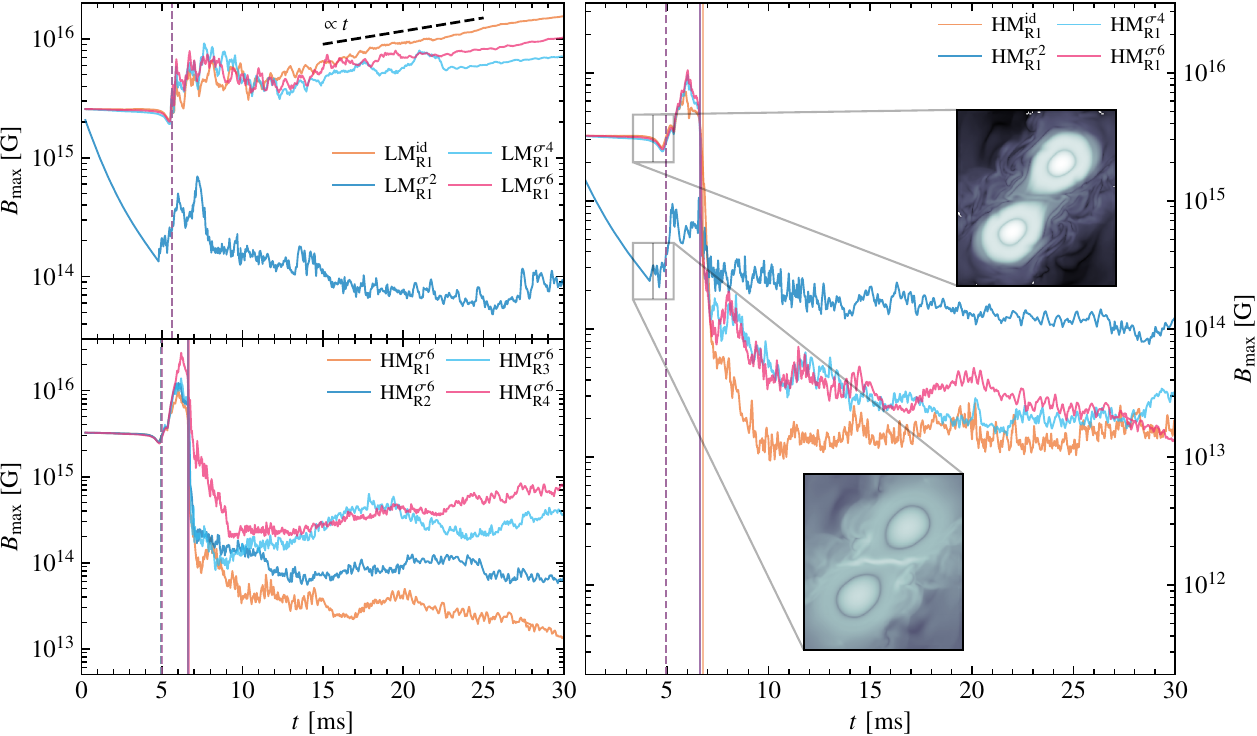}
    \caption{Maximum magnetic field for the runs described in
    \cref{tab:bns_sim}.
    In each panel, the dashed vertical lines and the solid vertical lines show
    the merger times and BH formation times, respectively. 
    The top left panel contains the LM simulations for varying $\sigma_0$ and
    the bottom left panel contains the HM simulations for $\sigma_0 = \num{e6}$
    at varying resolutions.
    The right panel shows the HM results for varying $\sigma_0$, with the two
    inset panels, each using the same color scale, showing the magnetic pressure
    for $\sigma_0 = \num{e6}$ (above) and $\sigma_0 = \num{e2}$ (below).
    For the former, the magnetic field remains primarily contained within the
    stars.
    For the latter, the magnetic pressure before the merger is essentially the
    same outside the stars as inside them.
    }
    \label{fig:bmax}
\end{figure*}

For comparing the magnetic field evolution in the different simulations, we
analyze its amplification during and after the merger for different
conductivities.
In \cref{fig:bmax}, we present the temporal evolution of the maximum magnetic
field strength.
In the top left panel, we compare its behavior for different conductivities in the
LM case when no BH is formed.
During merger, we observe a very quick amplification of the magnetic field,
which is maintained for the next $\sim$\qty{10}{ms}.
We associate this increase with the KHI, triggered by the shear layer at the
interface of the two colliding stars.
We note that our resolution is not high enough to fully resolve the turbulent
amplification process.
Subsequently, there is a slower continuous increase in the magnetic field
strength, most likely due to magnetic winding.
The values reach as high as \qty{e16}{\gauss} during our simulation time.
In the ideal GRMHD simulation, the increase is almost linear ($\propto t$),
which agrees with the expectations for magnetic winding.
For lower conductivities we obtain smaller slopes, while the maximum magnetic field
strength approaches the ideal case for increasing $\sigma_0$.
This behavior can be explained by considering that in the ideal approximation
the magnetic field is effectively ``frozen'' into the fluid.
Thus, magnetic winding and braking is more efficient in ideal GRMHD compared to
resistive GRMHD, where the fluid can move across magnetic field
lines~\cite{Dionysopoulou2015, Shibata2021}.
Notably, the case with $\sigma_0 = 10^2$ is significantly different from the
other simulations, as the neutron stars cannot sustain the initial magnetic
field in their interiors over the duration of the inspiral.
In fact, the maximum magnetic field strength inside the stars drops to
$\sim$\qty{e14}{\gauss} prior to merger.
This is a direct result of the short timescale of Ohmic diffusion for low
conductivities, as previously seen in \cref{sec:tov-pol} for the case of the
single neutron star.
There we observe that with $\sigma_0 = \num{e2}$, the magnetic field strength
decays by an order of magnitude after only \qty{3}{ms}, which is similar to
results seen here in the binary case.

In the right panel of \cref{fig:bmax} we show the maximum value of the
magnetic field for the HM simulations that form a BH a few milliseconds after
the merger.
Before the collapse of the remnant, the evolution is similar to the LM
simulations, where the magnetic field increases at merger. 
At BH formation, however, the magnetic field strength drops significantly, as
parts of the magnetic field are swallowed during the collapse of the remnant.
\qty{15}{ms} after BH formation, the maximum value of the magnetic field is
almost the same for the ideal case and the simulations with $\sigma_0 = 10^4$
and $\sigma_0 = 10^6$. 
For the low conductivity case with $\sigma_0 = 10^2$, the magnetic field starts
to oscillate even before the merger and roughly keeps this value even after the
BH formation, which is larger than the higher conductivity cases.
The electrodynamics are here so diffusive that the magnetic pressure outside the
star is comparable with the internal field (see inset panels of
\cref{fig:bmax}), which decays by about one order of magnitude pre-merger in
comparison to the higher conductivity and ideal simulations.

In the bottom left panel, we show the maximum of the magnetic field for the
$\sigma=10^6$ case for increasing resolution.
We find that for higher resolutions the magnetic field is amplified after the BH
formation.

\begin{figure*}[t]
\centering
    \includegraphics[width=\linewidth]{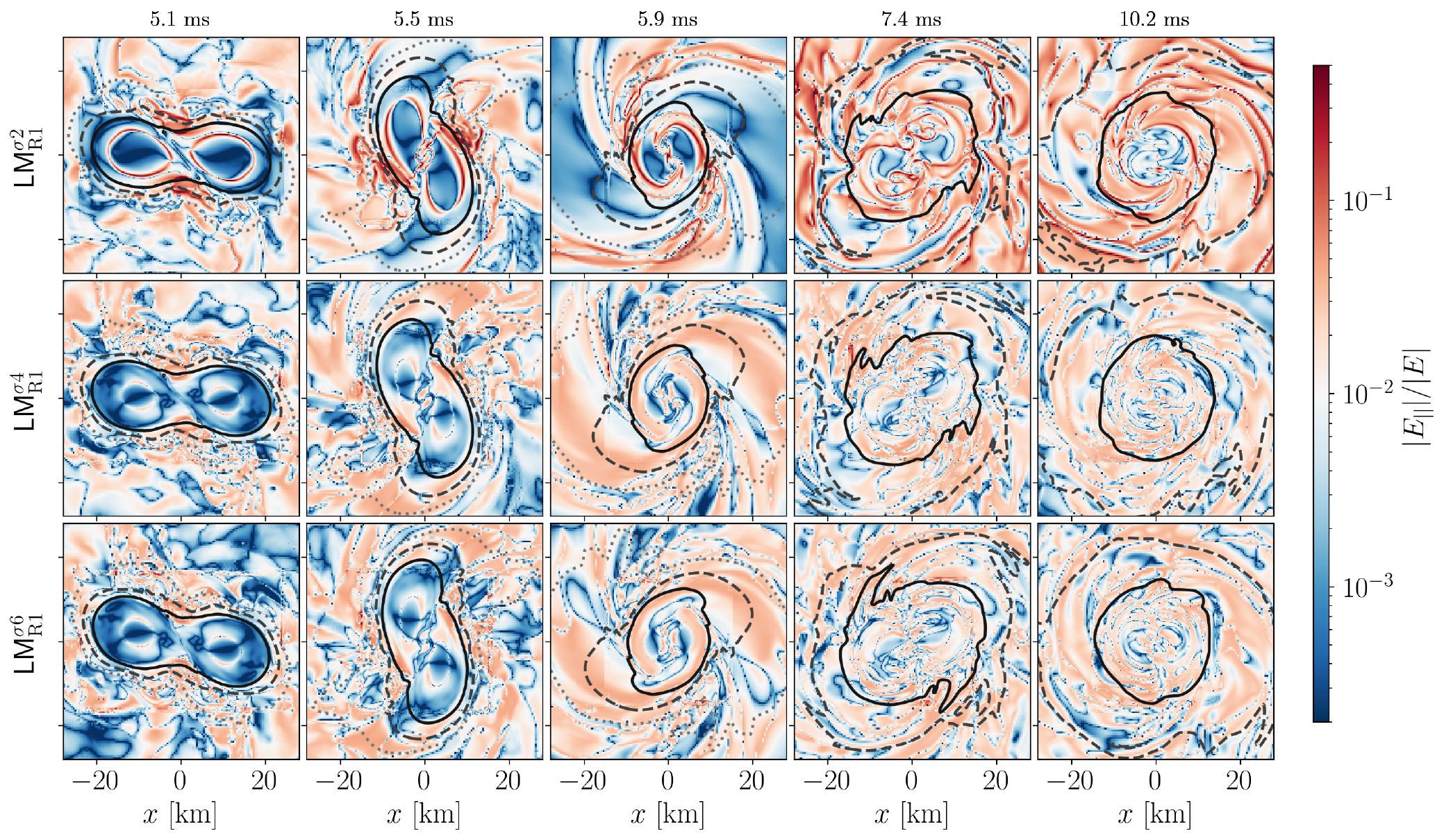}
    \caption{Relative electric field parallel to the magnetic field for the LM
    simulations with $\sigma_0 = \{10^2,10^4,10^6\}$ at the R1 resolution.
    Each row shows snapshots for one simulation in the $xy$-plane, while each
    column represents different times in the simulation denoted at the top.
    Additionally, isodensity contour lines are shown for $\rho = \{\num{e10},
    \num{5e11}, \num{e13}\}$~\unit{g.cm^{-3}} in gray to black scales as
    dotted, dashed, and solid lines, respectively.}
    \label{fig:BNS-Epara_LM}
\end{figure*}

\begin{figure*}[t]
    \centering
    \includegraphics[width=\linewidth]{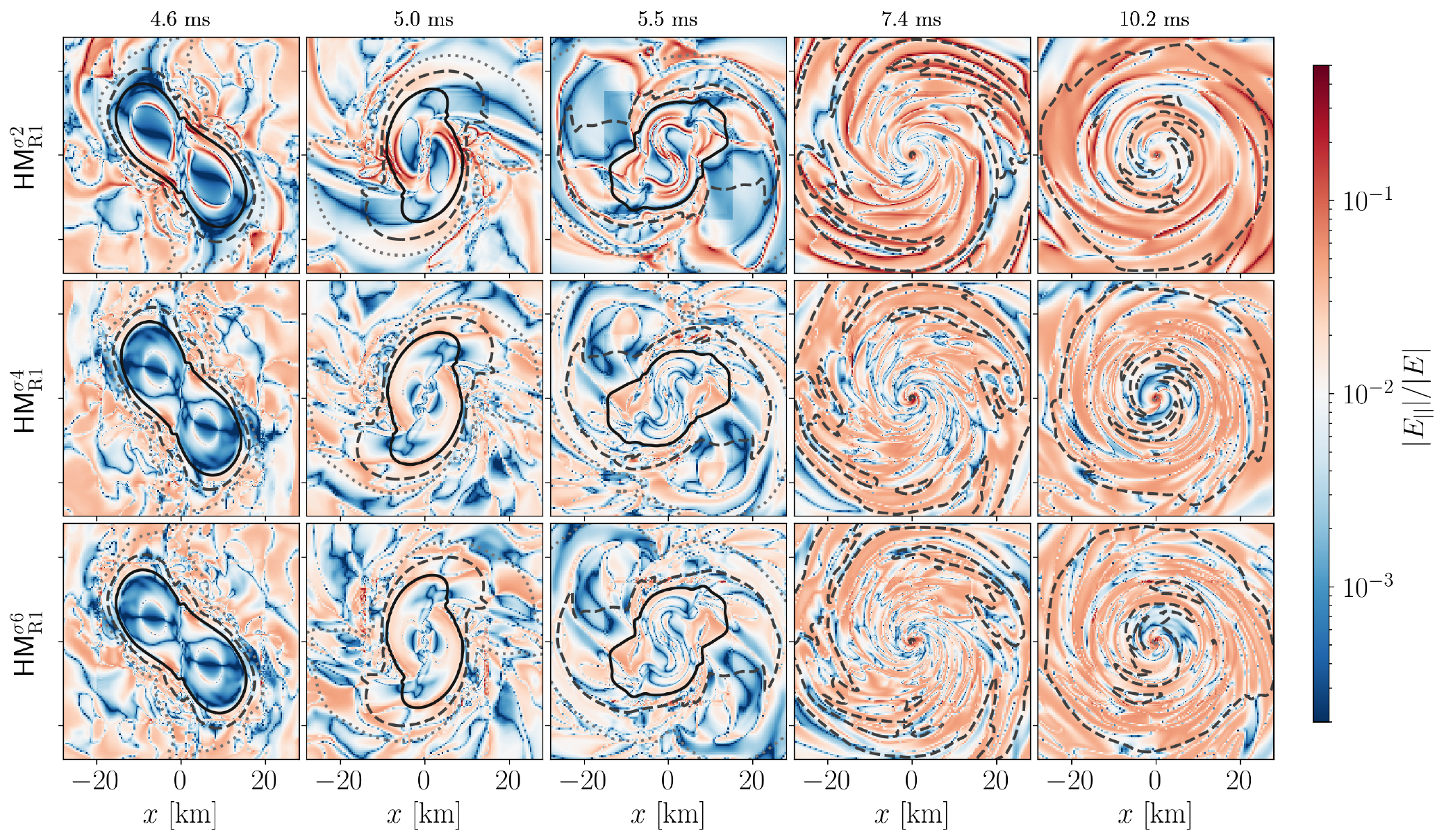}
    \caption{Relative electric field parallel to the magnetic field for the HM
    simulations with $\sigma_0 = \{10^2,10^4,10^6\}$ at the R1 resolution.
    Each row shows snapshots for one simulation in the $xy$-plane, while each
    column represents different times in the simulation denoted at the top.
    Additionally, isodensity contour lines are shown for $\rho = \{\num{e10},
    \num{5e11}, \num{e13}\}$~\unit{g.cm^{-3}} in gray to black scales as
    dotted, dashed, and solid lines, respectively.}
    \label{fig:BNS-Epara_HM}
\end{figure*}

To assess regions in which non-ideal GRMHD effects might become relevant in BNS
mergers, we consider the component of the electric field parallel to the
magnetic field
\begin{equation}
    E_\parallel = \frac{E_i B^i}{\sqrt{B_i B^i}}.
\end{equation}
In the ideal GRMHD limit, the electric field is perpendicular to the magnetic
field, implying $E_\parallel = 0$, whereas resistive GRMHD allows for non-zero
values of $E_\parallel$.
In Figs.~\ref{fig:BNS-Epara_LM} and \ref{fig:BNS-Epara_HM}, we show snapshots
of $E_\parallel$ relative to the total electric field strength $|E| = \sqrt{E_i
E^i}$ in the orbital plane for the LM and HM simulations, respectively.
As expected, the simulations with $\sigma = 10^{2}$ reveal the strongest
deviations from the ideal GRMHD behavior.
Nevertheless, even at such unrealistically low conductivities, $E_\parallel$
remains negligible inside the neutron star cores at the onset of merger.
The highest values are reached near the stellar surfaces and at the collision
interface during the merger.
For higher conductivities, the relative $E_\parallel$ component still stays
well below 10\%.
In the LM simulations, which form a HMNS remnant, the values inside the remnant
lie around $\sim$1\%.
Since realistic conductivities are expected to be even larger at the
corresponding high rest-mass densities of the remnant, these results suggest
that non-ideal GRMHD effects remain subdominant.
In contrast, HM simulations exhibit higher values for $E_\parallel / |E|$ when
the BH is formed.
In the surrounding disk with lower rest-mass densities, values of
$\lesssim$10\% are reached, indicating that in this scenario non-ideal GRMHD
effects become more relevant.

\subsection{Matter outflow and remnant properties}
\begin{figure}
    \centering
    \includegraphics[width=0.48\textwidth]{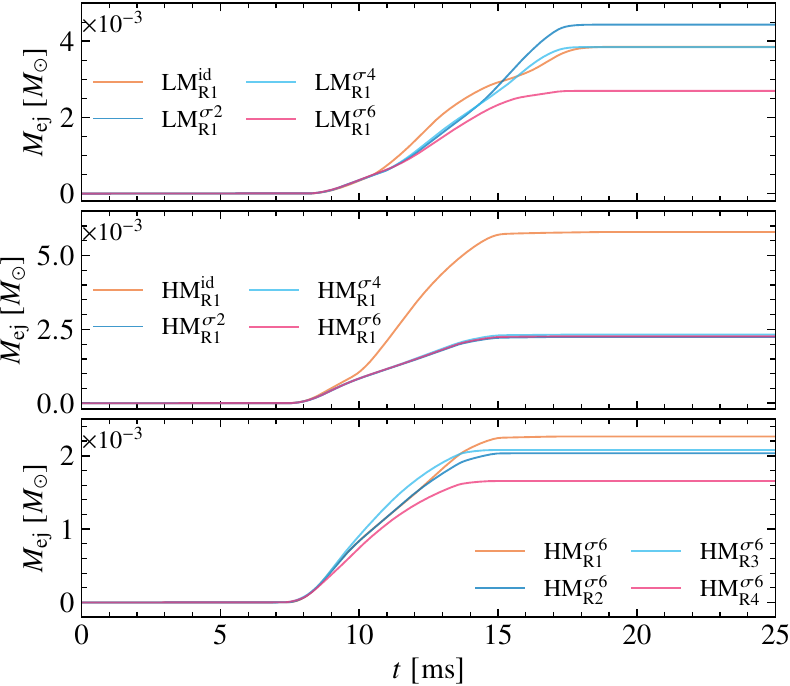}
    \caption{Unbound ejecta profiles at radius \qty{300}{\msun} for the runs
    described in \cref{tab:bns_sim}.
    The difference between the ideal case and the $\sigma_0 = 10^6$ case, in
    particular for the HM simulations, is due to differences in the evolution
    equations which slightly affect the lapse through the simulation, specially
    at the BH formation.}
    \label{fig:ejecta}
\end{figure}

To investigate the potential influence of the magnetic field description on
electromagnetic counterparts, we are examining simple ejecta properties such as
the ejecta mass.
A more detailed study of BNS merger outflows within a resistive GRMHD
description is left for future work.
We extract ejecta information using a coordinate sphere with radius
\qty{300}{\msun} centered around the origin.
We calculate on this sphere the amount of matter assumed to be unbound according
to the geodesic criterion~\cite{Hotokezaka2013}, \ie fulfilling $u_t < -1$ and
$u_r > 0$.

The main findings are presented in \cref{fig:ejecta}.
For the LM setups, shown in the upper panel, we find that the amount of ejecta,
generally of the order of \qty{3e-3}{\msun}, decreases with increasing
conductivity.
The ideal GRMHD code produces results closest to the $\sigma_0 = 10^4$ case and
similar to the $\sigma_0 = 10^6$ case.
Given the sensitivity of the ejecta mass to the exact numerical setup, we
attribute this variation primarily to implementation differences, \eg the
employed conservative-to-primitive routines, the timestepping procedure, or the
specific coupling of the electromagnetic fields to the spacetime. 
Regarding the latter, comparing the lapse between the resistive ($\sigma_0 =
10^6$) and the ideal case shows minor differences during the first \qty{5}{ms}
after merger, which can amount to slightly different mass outflows. 
Consequently, while we find that the trend of the ejecta mass seems to be
robust, \ie a decrease in ejecta mass for increasing conductivity, the
absolute value has to be assigned with an error of about 50\% to 100\%.
This becomes even more evident in the HM case, where a black hole forms.
Here, we find differences between the ideal and resistive GRMHD cases exceeding
100\%.

In the HM simulations (center panel of \cref{fig:ejecta}), the fast
collapse of the remnant into a BH during the first milliseconds after merger
leads to no measurable difference in the ejecta mass for different
conductivities, and most of the magnetic field energy is contained quickly
within the newly formed BH (\cf\cref{fig:bmax}).
Hence, all resistive HM simulations yield a very similar ejecta mass of about
\qty{2.5e-3}{\msun}.
On the contrary, the ejecta mass for the ideal case is slightly above
\qty{5e-3}{\msun}.
We attribute this to a slight difference in the implementation becoming
most apparent close to black hole formation.
For example, although both code versions lead to a BH formation at the same
time, their lapse functions at this time behave slightly differently.

In the lower panel of \cref{fig:ejecta}, we show the ejecta mass for the
$\sigma_0 = \num{e6}$ HM case for different resolutions.
Generally, we find a difference of about 20\% between resolutions R2 and R4,
with an overall decreasing trend as resolution increases. 

\section{Conclusion}\label{sec:conclusion}
We have implemented a new resistive general-relativistic magnetohydrodynamics
module as an extension to the existing GRMHD module~\cite{Neuweiler2024}
in \bam{}.
This extension enables us to incorporate non-ideal effects, such as physical
dissipation and diffusion.
Before, such effects were absent at the continuum limit of the evolved
equations, and only appeared artificially through the numerical discretization.
However, the added complexity comes with a price, namely additional
computational cost.
The electric field, which now must be evolved independently, is governed by an
evolution equation that becomes stiff for large electrical conductivities (\ie
the ideal limit).
We approach this issue by employing an IMEX Runge--Kutta method to efficiently
treat the non-stiff variables explicitly and implicitly evolve the electric
field.
The result is a unified framework for astrophysical plasmas with conductivities
spanning many orders of magnitude.
The implicit solve and conservative-to-primitive recovery, which are inherently
coupled in this formulation, are handled by a direct update of the electric field
enabled by the specific choice for Ohm's law and a recovery scheme following
the work of \citet{Palenzuela2009}.

Although we have so far only worked with an isotropic scalar version of Ohm's
law, the possibility exists within the code to modify the expression to include
additional physics such as mean-field dynamo terms (\eg\cite{Bucciantini2013})
or magnetic field--induced anisotropies~\cite{Palenzuela2013}.
The latter permits a transition from ideal to force-free electrodynamics (as
opposed to ideal to electrovacuum) via a phenomenological current, which would
change the dynamics of the EM fields in regions of lower rest-mass density, such
as the remnant disk and possible magnetosphere, and may affect the magnetic
structures in these regions or the outflow of EM energy via radiation.
Depending on the exact form of the expression for the current, a direct
(analytic) update of the electric field may not be possible; in that case, the
infrastructure already exists within \bam{} to numerically solve the implicit
update equation.

In order to validate our implementation, we performed nine special-relativistic
tests in one to three dimensions along with a small set of general-relativistic
simulations of a single, non-rotating, magnetized neutron star.
These tests demonstrated the robustness of the code for small and large uniform
conductivities as well as conductivities that vary several orders of magnitude
across the grid.
Where possible, the results were compared with exact solutions or the
corresponding result from \bam{}'s established and tested ideal GRMHD module,
and in all cases, the new module accurately recreated the exact solution,
matched the ideal result, or reproduced the existing results in the
literature.

Along with \cite{Palenzuela2013a, Palenzuela2013b, Ponce2014,
Dionysopoulou2015}, the simulations presented in this work complement the rather
limited number of BNS mergers performed with resistive GRMHD in the literature.
For these initial simulations, we compared the results obtained with ideal GRMHD
and those obtained using the new module. We saw minor differences between the
high-conductivity and ideal cases, which are attributable mainly to minor 
implementation differences.
By examining the component of the electric field parallel to the magnetic field,
which is identically zero in ideal GRMHD, we obtain a measure of deviation from
the ideal result in the resistive simulations.
For the high-conductivity case, we observed as expected negligible deviation
inside the neutron stars up the onset of merger.
However, the lower--rest-mass density disk surrounding the remnant black hole
carries a correspondingly lower conductivity, which is properly handled as a
transition region by the code.
The relative magnitude of the parallel component reaches values of $\leq$10\%. 
This provides further indication of the importance of non-ideal GRMHD effects in
the post-merger disk, and although not examined here, this connection can be
extrapolated to the important jet-launching process.
Future studies at higher resolution and a wider configuration space would be
required for a stronger quantitative statement on the scale of importance of
non-ideal effects.

\section*{Acknowledgments}
We thank B.~Br\"ugmann and J.~Kupka for useful discussions throughout the work
on this project.
R.J., H.G., and T.D.\ acknowledge funding from the EU Horizon under ERC Starting Grant,
no.\ SMArt-101076369. A.N.\ and T.D. gratefully acknowledge support from the
Deutsche Forschungsgemeinschaft, DFG, project number 504148597 (DI 2553/7).
M.U.\ acknowledges the S\~ao Paulo Research Foundation (FAPESP), Brasil, for
financial support under Process No.\ 2024/21086-5. 
Computations presented in this paper were run on the DFG-funded research cluster
Jarvis at the University of Potsdam (INST 336/173-1; project number:
502227537).
Views and opinions expressed are those of the authors only and do not
necessarily reflect those of the European Union or the European Research
Council.
Neither the European Union nor the granting authority can be held responsible
for them.

\bibliographystyle{apsrev4-2}
\bibliography{refs.bib}

\end{document}